\documentclass[aps,prb,twocolumn,showkeys,showpacs,amsmath,amsfonts,floatfix]{revtex4-1}
\usepackage[T1]{fontenc}
\usepackage{graphicx}
\usepackage[caption=false]{subfig} 
\usepackage{tikz} 
\usepackage[colorlinks=true,citecolor=blue,linkcolor=blue,urlcolor=blue]{hyperref}%
\def\be{\begin{equation}}
\def\ee{\end{equation}}
\def\bea{\begin{eqnarray}}          
\def\eea{\end{eqnarray}}
\def\bi{\begin{itemize}}
\def\ei{\end{itemize}}

\newcommand{\bl}[1]{{\color{blue}{{#1}}}}
\newcommand{\gn}[1]{{\color{green}{{#1}}}}

\begin{document}
\title{    
Dynamics of a quantum phase transition in the 1D Bose-Hubbard model:\\
excitations and correlations induced by a quench
}
\author{Bart{\l}omiej Gardas}
\affiliation{Theoretical Division, LANL, Los Alamos, New Mexico 87545, USA}             
\affiliation{Institute of Physics, University of Silesia, 40-007 Katowice, Poland}     
\affiliation{Instytut Fizyki Uniwersytetu Jagiello\'nskiego,
             ulica {\L}ojasiewicza 11, PL-30-348 Krak\'ow, Poland}        
\author{Jacek Dziarmaga} 
\affiliation{Instytut Fizyki Uniwersytetu Jagiello\'nskiego,
             ulica {\L}ojasiewicza 11, PL-30-348 Krak\'ow, Poland}
\author{Wojciech H. Zurek}
\affiliation{Theoretical Division, LANL, Los Alamos, New Mexico 87545, USA}             
\date{\today}
\begin{abstract}
The ground state of the one-dimensional Bose-Hubbard model at unit filling undergoes the Mott-superfluid quantum phase transition. It belongs to the Kosterlitz-Thouless universality class with an exponential divergence of the correlation length in place of the usual power law. We present numerical simulations of a linear quench both from the Mott insulator to superfluid and back. The results satisfy the scaling hypothesis that follows from the Kibble-Zurek mechanism (KZM). In the superfluid-to-Mott quenches there is no significant excitation in the superfluid phase despite its gaplessness. Since all critical superfluid ground states are qualitatively similar, the excitation begins to build up only after crossing the critical point when the ground state begins to change fundamentally. The last process falls into the KZM framework.
\end{abstract}
\maketitle

\section{Introduction}
\label{sec:intro}

The study of the dynamics of phase transitions started with the question posed by Kibble \cite{Kibble76,Kibble80}. He noted that, in the rapidly cooling post -- Big Bang Universe, phase transitions must lead to disparate local choices of the broken symmetry vacuum. The resulting mosaic of domains with independently chosen vacua will in turn precipitate formation of topological defects with observable consequences. For instance, Kibble theorized that the presently observed cosmological structures have been seeded by cosmic strings -- an example of such defects. 

While the original estimate \cite{Kibble76} of defect density was incorrect, as it relied on an equilibrium argument \cite{Hindmarsh17} (hence, densities did not depend on the phase transition rate), the key question -- whether transitions leave relics such as topological defects in their wake -- can be also posed in the condensed matter setting and studied in the laboratory. The theory \cite{Zurek85,Zurek93,Zurek96} developed to estimate defect density for the second order phase transition relies on their non-equilibrium nature but also on the universality class of the transition: It uses equilibrium critical exponents to predict scaling of the density of defects and other excitations with the quench timescale. The size of domains within which symmetry breaking can be coordinated is given by the size of the ``sonic horizon'' that estimates how far the local choice of the broken symmetry can be communicated (and plays a similar role as the causal horizon used byKibble \cite{Kibble80} to set lower bounds on defect density)

The usual estimate of the sonic horizon size relies on a power-law scaling of the relaxation time and the healing length 
with a distance from the critical point that is characterized by the dynamical and correlation length critical exponents $z$
and $\nu$. A characteristic time-scale 
$
\hat t \sim \tau_Q^{z\nu/(1+z\nu)}
$
and a length-scale
$
\hat\xi \sim \tau_Q^{\nu/(1+z\nu)}
$
are predicted, where the quench time $\tau_Q$ quantifies the rate of the transition. The correlation length $\hat\xi$ 
determines the number of excitations as a function of $\tau_Q$. The Kibble-Zurek mechanism (KZM) has been confirmed by numerical simulations \cite{LagunaZ1,YZ,DLZ99,ABZ99,ABZ00,BZDA00,ions20,ions2,WDGR11,dkzm1,dkzm2,Nigmatullin11,DSZ12,holo2,Sonner,Francuzetal} 
and by experiments \cite{Chuang91,Bowick94,Ruutu96,Bauerle96,Carmi00,Monaco02,Monaco09,Maniv03,Sadler06,Golubchik10,Chae12,Griffin12,
Schaetz13,EH13,Ulm13,Tanja13,Anderson08,Lamporesi13} in a variety of settings, with most recent results providing evidence of the KZM scaling laws \cite{Chae12,Griffin12,DalibardSupercurrents,DalibardCoherence,ferroelectrics,Hadzibabic,Navon,FerroKZscaling,Chicago}. 
Refinements and extensions of KZM include phase transition in inhomogeneous systems (see~\onlinecite{DKZ13} for a recent overview),
generation of winding numbers in the limit when flux trapping is a rare event \cite{Monaco09}
(and KZM scaling must be suitably adjusted \cite{Zurek13}),
and applications that go beyond topological defect creation (see e.g.~\onlinecite{DQZ11,Zurek09,DZ10,Cincio}). Recent reviews related to KZM are also available~\cite{Kibble03,Kibble07,Dziarmaga10,Polkovnikov11,DZ13}.

In this paper we consider a zero-temperature quantum phase transition. Despite important differences with respect to thermodynamic transitions -- where thermal rather than quantum fluctuations act as seeds of symmetry breaking -- the KZM can be generalized to quantum phase transitions
\cite{Bishop,Damski2005,Dorner2005,Dziarmaga2005,Polkovnikov2005,ind1,ind2,ind3,ind4,ind5,ind6,ind7,ind8,ind9,ind10,ind11,ind12,ind13,Francuzetal}, 
see also~\onlinecite{Dziarmaga10,Polkovnikov11,DZ13} for reviews.
The quantum regime was also addressed in some of the recent experiments~\cite{Esslinger,deMarcoclean,Schaetz,FerroKZscaling,deMarcodisorder,chinskiLZ,Chicago}.

Recently a scaling hypothesis that involves both space and time was proposed \cite{Kolodrubetz,princeton,Francuzetal} as a generalization and extension of the predictive power of KZM, though some of its basic ingredients were known from the beginning~\cite{Zurek96,Cincio,ViolaOrtiz,DamskiZurek,DzRams}. Since in the adiabatic limit, when $\tau_Q\to\infty$, both scales $\hat t$ and $\hat \xi$ diverge, they should become the only relevant time and length scales in the regime of low frequencies and long wavelengths. This in turn suggests a dynamical scaling hypothesis, similar to the (static) one in equilibrium phase transitions, that during the quench all physical observables depend on a time $t$ through the scaled time $t/\hat t$ and on a distance $x$ through the scaled distance $x/\hat\xi$. What makes it really powerfull is the fact that the two scales are not independent: $\hat t\sim\hat\xi^z$. This space-time renormalization hypothesis was confirmed in a precise experiment~\cite{Chicago} where it proved useful in organizing the experimental data. In the following we also find it useful in organizing our numerical results.

In this paper we consider the Mott-superfluid quantum phase transition in the 1D Bose-Hubbard model that belongs to the Kosterlitz-Thouless universality class. This problem was touched upon in Ref. \onlinecite{KZSR} where it was argued that, since the correlation length diverges exponentially near the critical point, one cannot ascribe a definite scaling exponent $w$ to $\hat\xi\sim\tau_Q^w$ except for $w=1$ in the limit of exceedingly slow $\tau_Q$ that are beyond any realistic experiment. 
However, when the range of $\tau_Q$ is restricted to one or two orders of magnitude, then an effective scaling 
$\hat\xi\sim\tau_Q^w$ with an effective exponent $w<1$ can be a convenient approximation. In this paper, we readdress the problem, this time with fully fledged numerical DMRG simulations. Like in a real experiment, there are limitations that restrict the range of available quench times and the KZ scaling hypothesis with effective exponents is a convenient approximation. We simulate linear quenches from the Mott insulator to superfluid, where the range of correlations builds up as the tunnelling rate between nearest-neighbor sites is turned on. The spatial profile of the correlators and their time dependence satisfy the scaling hypothesis. 

We also simulate reverse linear quenches from the superfluid to Mott insulator. Somewhat surprisingly, we find the excitation during a ramp across the gapless superfluid to be negligible as compared to the excitation that builds up after crossing the critical point to the Mott phase. Apparently, the different critical superfluid ground states that are crossed by the linear ramp are similar enough for the excitation to be negligible despite their vanishing gap. What matters here is that a relatively large change of the tunneling rate during the ramp across the superfluid phase corresponds to a relatively small change of the Luttinger liquid parameter $K$ that determines the ground state of the liquid. When measured by a distance between different ground states, the superfluid phase can be effectively identified as a single critical point. 
It is only after crossing to the Mott phase that the ground state begins to change fundamentally. This inevitably makes the excitations build up until their growth is halted by the opening Mott gap. The last crossover, that takes place at the time $\hat t$ after crossing to the Mott phase, is the essence of the quantum KZM. Our simulations confirm this simple scenario by demonstrating that the excitation energy in the Mott phase satisfies the KZ scaling hypothesis.

The paper is organized as follows. We begin with a general discussion of the textbook version of the quantum Kibble-Zurek mechanism in section~\ref{sec:QKZ}. This generic version assumes gapfull phases on both sides of the transition and a power law divergence of the correlation length at the critical point. These assumptions are relaxed in section~\ref{sec:KZinKT} where we introduce the 1D Bose-Hubbard model and consider the phase transtion from Mott insulator to superfluid that belongs to the Kosterlitz-Thouless universality class. Here the divergence of the correlation length on the Mott side is exponential and the whole superfluid phase is gapless and critical. We recall relevant predictions of Ref.~\onlinecite{KZSR} for this type of transition. In this paper these results are corroborated by numerical simulations described in sections~\ref{sec:MISF} and~\ref{sec:SFMI}. In section \ref{sec:MISF} we describe simulations of the Mott-to-superfluid transition and in section~\ref{sec:SFMI} those of the reverse superfluid-to-Mott quench. Finally, we briefly conclude in section~\ref{sec:concl}.

\section{ Quantum Kibble-Zurek mechanism } \label{sec:QKZ}

%
\begin{figure}[htp!]
	\begin{tikzpicture}[scale=1.0] 
	\draw [ultra thick, ->] (-4,0)--(4,0);	
	\draw [ultra thick, ->] (0,0)--(0,4.);
	\node [right] at (4,0) {\large $t$};
	\draw [thick] (-1.5,-.1) node[below]{\large $-\hat{t}$} -- (-1.5,0.1);
	\draw [thick] (0,-.1)  node[below]{\large $0$}  	  -- (0,0.1);
	\draw [thick] (1.5,-.1)  node[below]{\large $+\hat{t}$} -- (1.5,0.1);
	\node at (-3,3) {adiabatic};
	\node at (+3,3) {adiabatic};
	\node at (0,4.3) {impulse};
	%
	%
	\node[scale=1.6] at (-3.7,0.8) {\small \bl{rate}};
	\node[scale=1.6] at (+3.5,1.75) {\small \gn{gap}};
	\draw[scale=0.6,domain=-6:-0.6,smooth,variable=\x,blue] plot ({\x},{-4./\x});
	\draw[scale=0.6,domain=0.6:6,smooth,variable=\x,blue] plot ({\x},{+4./\x});
	\draw[scale=0.6,domain=0.:2.5,smooth,variable=\y,green] plot ({\y*\y},{\y});
	\draw[scale=0.6,domain=0.:2.5,smooth,variable=\y,green] plot ({-\y*\y},{\y});		
	\draw [ultra thick,dashed,red] (-1.5,0) -- (-1.5,4);
	\draw [ultra thick,dashed,red]  (1.5,0) -- (1.5,4);
	\fill[gray!20,nearly transparent] (-1.5,0) -- (1.5,0) -- (1.5,4) -- (-1.5,4) -- cycle;
	\end{tikzpicture}
	\caption{\label{fig:KZM1}
       As a system is driven across a generic quantum critical point with a linear ramp $\epsilon(t)=t/\tau_Q$, 
       the energy gap between the ground state and the first relevant excited state closes like $|\epsilon|^{z\nu}$ and, at the same time, the transition rate diverges like 
       $1/|t|$. The two are equal at $\pm\hat t$, where $\hat t \sim \tau_Q^{z\nu/(1+z\nu)}$. 
       The evolution must be non-adiabatic between $-\hat t$ and $\hat t$.
	}	
\end{figure}
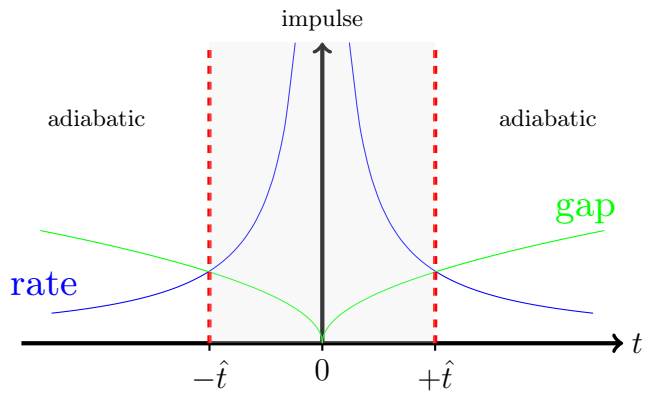

A distance from a quantum critical point can be measured with a dimensionless parameter $\epsilon$. The ground state of the 
Hamiltonian $H(\epsilon)$ undergoes a fundamental change at $\epsilon=0$ when the correlation length in its ground state 
diverges like
\be 
\xi \sim |\epsilon|^{-\nu}
\label{xi}
\ee
and the relevant gap closes like
\be 
\Delta \sim |\epsilon|^{z\nu}.
\label{Delta}
\ee
The system, initially prepared in its ground state, is driven across the critical point by a linear quench,
\be 
\epsilon(t)=\frac{t}{\tau_Q},
\label{quench}
\ee 
with a quench time $\tau_Q$. 

The evolution sufficiently far from the critical point is initially adiabatic. However, the rate of change of epsilon,
\be 
\left|\frac{\dot\epsilon}{\epsilon}\right|=\frac{1}{|t|}, 
\label{rate}
\ee
diverges at the gapless critical point. Therefore, the evolution cannot be adiabatic in its neighborhood between 
$-\hat t$ and $\hat t$, see Fig.~\ref{fig:KZM1}. Here $\hat t$ is the time when the gap~(\ref{Delta}) equals the 
rate~(\ref{rate}), so that:
\be 
\hat t \sim \tau_Q^{z\nu/(1+z\nu)}.
\label{hatt}
\ee
Just before the adiabatic-to-non-adiabatic crossover at $-\hat t$, the state of the system is still approximately the adiabatic 
ground state at $\epsilon=-\hat\epsilon$, where
\be 
\hat\epsilon = \frac{\hat t}{\tau_Q} \simeq \tau_Q^{-1/(1+z\nu)},
\label{hatepsilon}
\ee
with a correlation length
\be 
\hat\xi \sim  \hat\epsilon^{-\nu} \sim \tau_Q^{\nu/(1+z\nu)}.
\label{hatxi}
\ee
In a first-order impulse approximation, this state ``freezes out'' at $-\hat t$ and does not change until $\hat t$. At $\hat t$ the frozen state is no longer the ground state but an excited state with a correlation length $\hat\xi$. It is the initial state for the adiabatic process that follows after $\hat t$.

No matter how quantitatively accurate is the above ``freeze-out scenario'', this simple scaling argument establishes $\hat\xi$ and $\hat t$, interrelated via 
\be 
\hat t\sim\hat\xi^z,
\label{hattz}
\ee 
as the relevant scales of length and time, respectively. What is more, in the adiabatic limit, when $\tau_Q\to\infty$, both scales diverge becoming the unique scales in the regime of long-wavelength and low-frequency. In analogy to static critical phenomena, this uniqueness suggests a scaling hypothesis \cite{Kolodrubetz,princeton,Francuzetal}:
\be 
\langle\psi(t)| O(x) |\psi(t)\rangle = \hat\xi^{-\Delta_O} F_O\left(t/\hat\xi^z,x/\hat\xi\right).
\label{Oscaling}
\ee
Here $|\psi(t)\rangle$ is the quantum state during the quench, $O(x)$ is an operator depending on a distance 
$x$, $\Delta_O$ is its scaling dimension, and $F_O$ its scaling function. The same conclusions about scaling are 
reached when one follows the narrative based on the sonic horizon \cite{Zurek96}.

\begin{figure}[htp!]
\begin{center}\includegraphics[width=1\columnwidth]{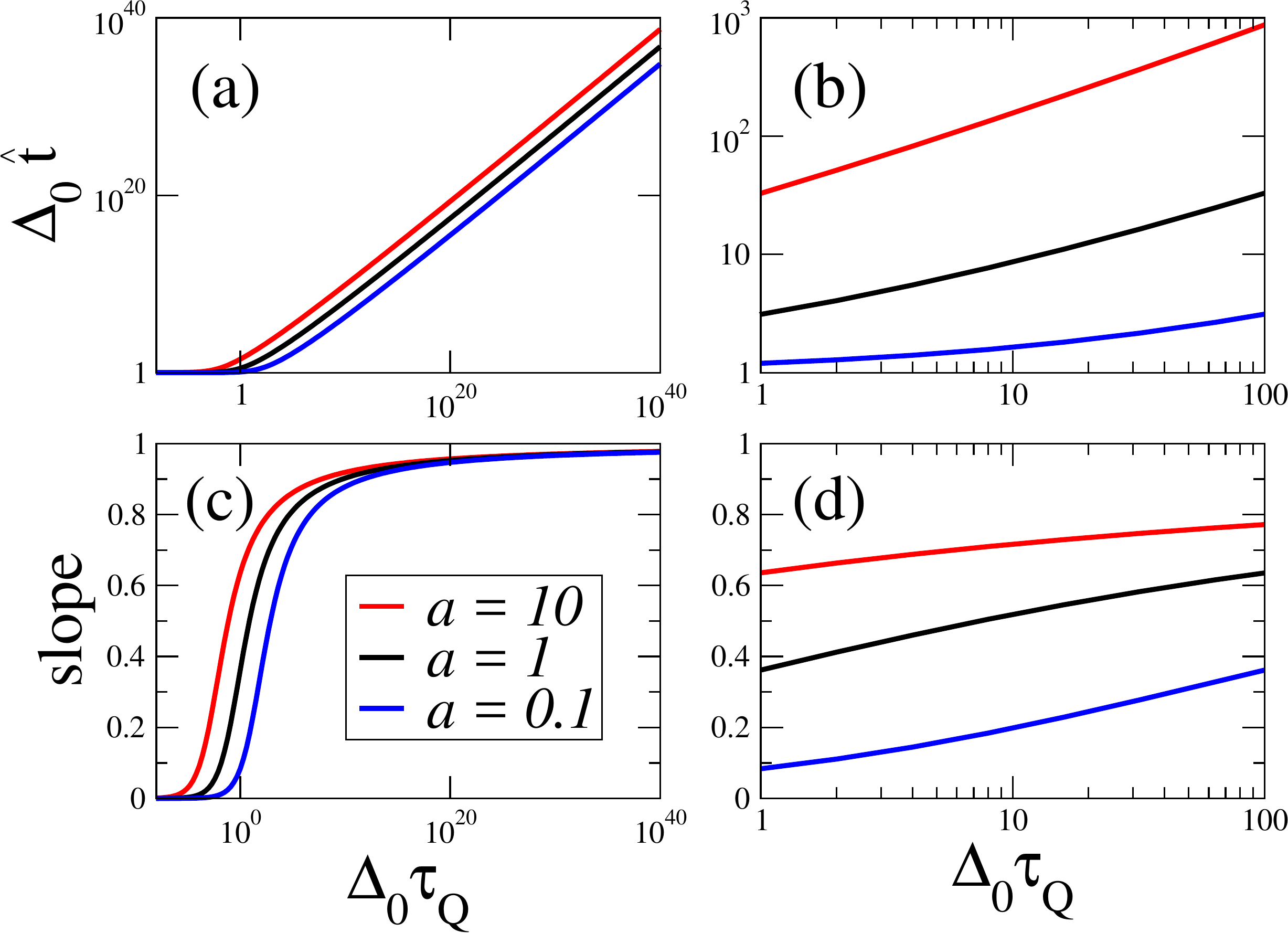}\end{center} 
\vspace{-0.0cm}
\caption{
(a) Log-log scale a generic KZ power-law $\hat t\propto\tau_Q^{\nu z/(1+\nu z)}$ in (\ref{hatt}) becomes a function
$
\ln_{10}\Delta_0\hat t = \frac{\nu z}{1+\nu z} \ln_{10}\Delta_0\tau_Q + {\rm const}
$,
where $\Delta_0$ is a microscopic energy scale. For comparison, in (a), we plot $\hat t$ for the Kosterlitz-Thouless 
transition in the 1D Bose-Hubbard model~(\ref{hattKT}) in function of $\tau_Q$ over many decades of the argument. 
This function becomes linear only asymptotically for $\tau_Q\to\infty$, but it may appear linear locally, i.e., 
in a range of one or two decades. Indeed, in (b), we focus on the narrow range of $\Delta_0\tau_Q=10^{0...2}$ that
is small enough e.g. for a realistic cold-atom experiment. These plots can be reasonably approximated by linear 
functions, especially when experimental error bars are present. In (c), we plot a local slope 
$
d\ln_{10}\left(\Delta_0\hat t\right)/d\ln_{10}\left(\Delta_0\tau_Q\right)
$ 
of the log-log plot in panel (a) in function of $\Delta_0\tau_Q$. The slope $1$ in Eq.~(\ref{hatxinaive}) is achieved asymptotically but only for $\tau_Q$ that are unrealistically large and imply correlations over distances that are ``astronomical'' in magnitude. In (d), a focus on the realistic $\tau_Q$ shows that the local slope can be significantly less than $1$.
}
\label{fig:hatt} 
\end{figure}

\section{KZ mechanism in the Kosterlitz-Thouless transition}
\label{sec:KZinKT}

The power laws~(\ref{xi},\ref{Delta}) are not directly applicable in the Kosterlitz-Thouless (KT) transition~\cite{KT1,KT2,KT3}, where, on the disordered/Mott side of the transition, the correlation length's divergence is exponential: 
\begin{equation}
\xi=\xi_0 \exp(2a/\sqrt{|\epsilon|}).
\label{xiKT}
\end{equation} 
Here $a\simeq 1$ and $\xi_0$ is a microscopic scale of length. This faster-than-polynomial divergence can be captured by stating that $\nu=\infty$, see \emph{e.g.}~\onlinecite{Girvin}, but it may tempt one to misuse Eqs.~(\ref{hatt},~\ref{hatepsilon},~\ref{hatxi}) by inserting $\nu=\infty$ together with $z=1$ to obtain:
\be 
\hat t \sim\tau_Q^1, \quad
\hat\epsilon\sim\tau_Q^0, \quad
\hat\xi\sim\tau_Q^1.
\label{hatxinaive}
\ee
As shown in Ref.~\onlinecite{KZSR}, these equations are valid asymptotically for $\tau_Q\to\infty$ but this 
asymptote is achieved for unrealistically slow $\tau_Q$ (and, hence, astronomically large $\hat\xi$ of the 
order of kilometers \cite{KZSR}). Below we briefly recount the argument.

In the 1D Bose-Hubbard model at commensurate filling, where $z=1$, the gap $\Delta\sim\xi^{-z}$ on the 
Mott-insulator side of the transition closes like
\begin{equation}
\Delta=\Delta_0 \exp(-2a/\sqrt{|\epsilon|}),
\label{xiKT2}
\end{equation} 
where $\Delta_0$ is a microscopic energy scale. For the linear ramp~(\ref{quench}), driving a quench 
from the Mott insulator to superfluid, this gap equals the rate~(\ref{rate}) at $t=-\hat t$ when
\begin{equation}
\Delta_0 \exp(-2a/\sqrt{\hat t/\tau_Q})=1/\hat t.
\end{equation} 
A solution of this transcendental equation is
\begin{equation} 
\hat t=
\tau_Q~
\frac{a^2}{\mathcal{W}^2\left(a\sqrt{\Delta_0\tau_Q}\right)},
\label{hattKT}
\end{equation}
where $\mathcal{W}$ is the Lambert function~\cite{Corless93}. The above solution is plotted for different values of $a\simeq1$ in Figs.~\ref{fig:hatt}a,b. A similar relation has been derived and confirmed by numerical 
simulations in a dissipative classical model~\cite{BKTclass}. It has been also tested in a recent experiment \cite{Deutschlander15}.

Figure~\ref{fig:hatt} shows that the exponent of unity for the dependence of $\hat t$ on $\tau_Q$ in 
Eq.~(\ref{hatxinaive}) is attained only for exceedingly slow quenches that are unlikely to be experimentally
or numerically accessible. For any reasonably slow quenches the effective exponent would be significantly less
than $1$.

The above argument determines the correlation length $\hat\xi$ imprinted on the quantum state on the Mott side 
of the transition. It characterizes the excited state when the quench ramp enters the superfluid phase.
In this critical phase the imprinted correlations are spreading with a velocity limited by the speed of 
quasiparticles.

\section{Quench from Mott insulator to superfluid}
\label{sec:MISF}

\begin{figure}[htp!]
\vspace{-0cm}
\subfloat{
\includegraphics[width=1.0\columnwidth,clip=true]{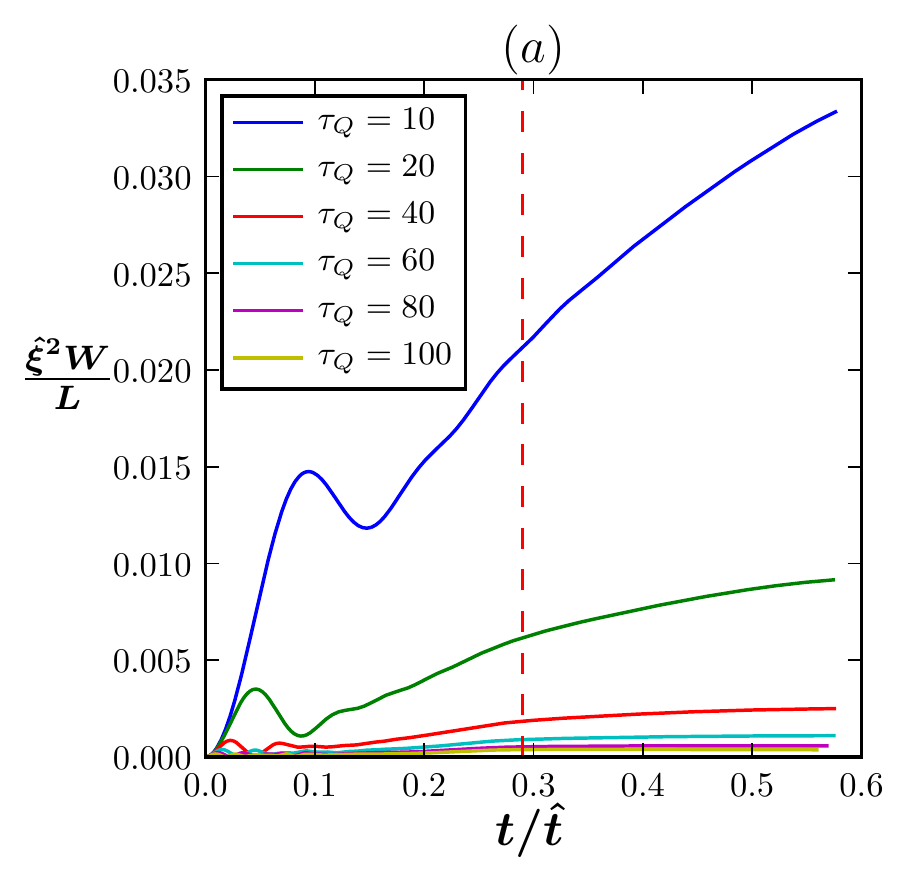}
\label{fig:DMRGvBOGa}
}
\newline
\subfloat{
\includegraphics[width=1.0\columnwidth,clip=true]{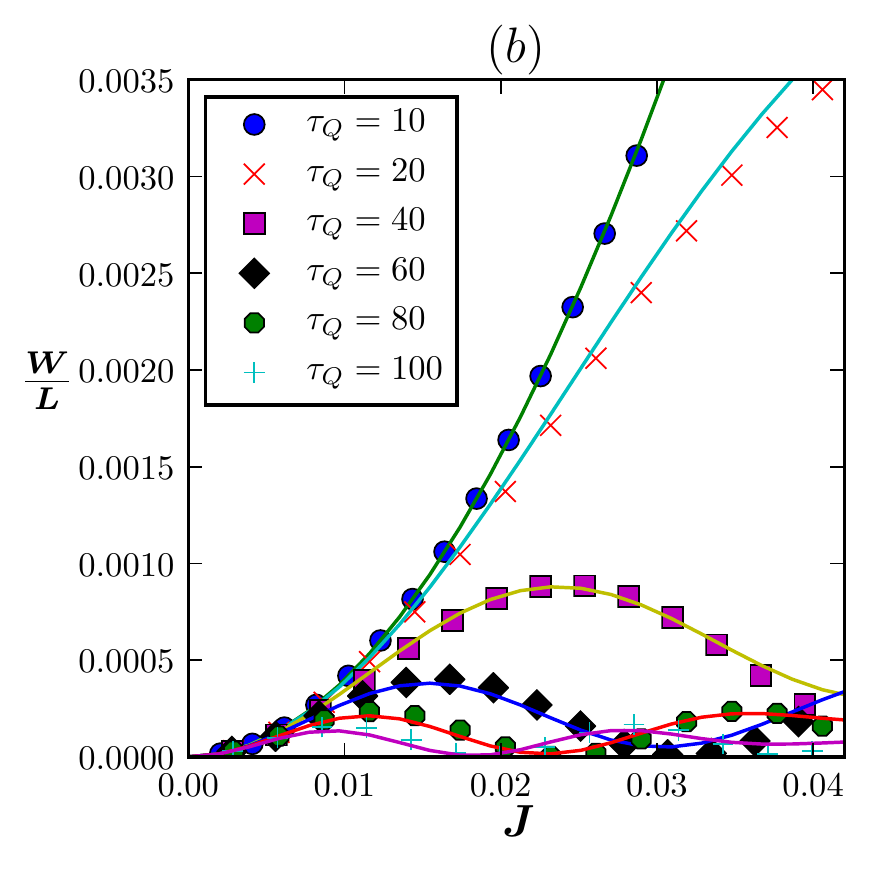}
\label{fig:DMRGvBOGb}
}
\vspace{-0cm}
\caption{
(a) Density of excitation energy $W/L$ in the center of $L=100$ sites in function of the hopping frequency $J$ - results from DMRG~\cite{Wall12} simulations for different $\tau_Q$. The early excitation visible at small $J$ originates from the discontinuos slope of the ramp~(\ref{eq:Jt}) at $J=0$. This initial excitation is suppressed adiabatically when $J$ is getting closer to $J_c=0.29$ (the vertical dashed line) and the gap is closing. The following excitation that begins to grow before $J_c$ is attributed to the KZ mechanism. For $\tau_Q=10$ the initial excitation is too strong and the KZ excitation begins too early for the two mechanisms to be clearly separated in $J$. 
(b) Focus on small $J$ where DMRG (solid lines) can be compared with the Bogoliubov theory for doublons and holons~\cite{DHBog} (data points). The agreement is better for faster $\tau_Q$ where the state remains closer to the initial Fock state~(\ref{Fock}) and has lower density of doublons and holons.
}
\label{fig:DMRGvBOG}
\end{figure}

The 1D Bose-Hubbard model at the commensurate filling of $1$ particle per site is described by a Hamiltonian
\be 
H=
-J\sum_{s=1}^{L-1} \left(b_l^\dag b_{l+1}+b_{l+1}^\dag b_l\right)+
\frac{U}{2}\sum_{l=1}^L n_l(n_l-1),
\ee
where $J$ is the hopping rate and $U$ is the on-site interaction strength.
The Kosterlitz-Thouless quantum phase transition from the Mott-insulator to superfluid takes place at 
$J_c=0.29$ (see \emph{e.g.} Ref.~\onlinecite{Jc}). We consider a quench driven by a linear ramp
\be 
J(t)=J_c
\left\{
\begin{array}{ll}
0,          & {\rm for~} t\leq-\tau_Q,\\
1+t/\tau_Q, & {\rm for~} t >  -\tau_Q.
\end{array}
\right.
\label{eq:Jt}
\ee
from $J(-\tau_Q)=0$ to $J(\tau_Q)=2J_c$. The initial state is the ground state at $J=0$:
\be 
|\psi(0)\rangle=|1,1,1,...,1\rangle.
\label{Fock}
\ee
Figure~\ref{fig:DMRGvBOG}\subref{fig:DMRGvBOGa} shows that there is a finite density of excitation energy $W/L$ above the adiabatic ground state during a quench, hence the evolution with the ramp is not adiabatic. Furthermore, the system seems to be excited in two stages 
by two different mechanisms.

\begin{figure}[htp!]
	\begin{tikzpicture}[scale=1.] 
	\draw [ultra thick, ->] (-4,0)--(4,0);	
	\draw [ultra thick, ->] (0,0)--(0,4.);
	\node [right] at (4,0) {\large $t$};
	\draw [thick] (-1.5,-.0) node[below]{\large $-\hat{t}$} -- (-1.5,0.1);
	\draw [thick] (0,-.0)  node[below]{\large $0$}  	  -- (0,0.1);
	%
	\node at (-3,3) {adiabatic};
	\node at (0,4.3) {impulse};
	%
	%
	\node[scale=1.6] at (-2.7,0.3) {\small \bl{rate}};
	\node[scale=1.6] at (-3.0,1.2) {\small \gn{gap}};
	\draw[scale=0.6,domain=-6:-0.57,smooth,variable=\x,blue] plot ({\x},{-3.8/\x});
	%
	\draw[scale=0.6,domain=-6:-0.001,smooth,variable=\x,green] plot ({\x},{2*exp(-0.5/sqrt(-\x))});
	\draw[scale=0.6,domain=0:6,smooth,variable=\x,green] plot ({\x},{0.1});		
	\draw [ultra thick,dashed,red] (-1.5,0) -- (-1.5,4);
	%
	\fill[gray!20,nearly transparent] (-1.5,0) -- (3.8,0) -- (3.8,4) -- (-1.5,4) -- cycle;
	\end{tikzpicture}
	\caption{\label{fig:KZM3}
    Quench from the Mott-insulator to superfluid, driven by the linear ramp (\ref{eq:Jt}), 
    the initial evolution before $-\hat t$ is adiabatic, then it becomes approximately impulse.
	}
	\label{fig:KZM2}
\end{figure}
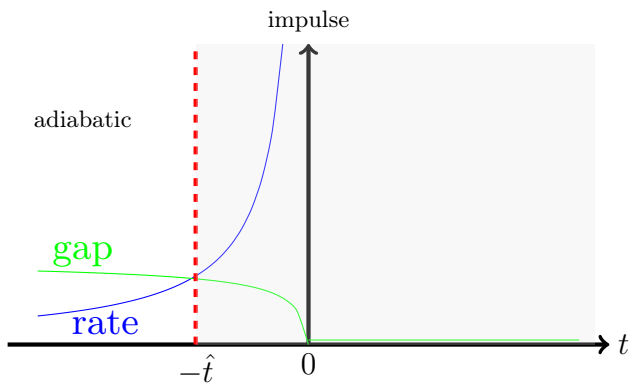

The early oscillations visible at small $J$ are excited by the discontinuous time derivative of the ramp at $J=0$ that is proportional to $1/\tau_Q$. Since this sudden initial jolt excites states well above the Mott gap with a probability proportional to $\tau_Q^{-2}$, the amplitude of the early excitation is roughly $W/L\propto\tau_Q^{-2}$, as can be demonstrated by the quantum perturbation theory \cite{Gritsev10}. For small enough $J$ they compare well with predictions of the Bogoliubov doublon-holon model~\cite{DHBog}, see figure~\ref{fig:DMRGvBOG}\subref{fig:DMRGvBOGb}. For larger $J$, as their energy gap closes with $J$ approaching $J_c$, the early oscillations are adiabatically suppressed. Closer to $J_c$ the KZ mechanism steps in, see Fig.~\ref{fig:KZM2}.

At the earlier stages of the project we attempted to mitigate the effect of the initial jolt by initiating the quench more smoothly. For rapid quenches such efforts proved only party successful. However, similarly as for the linear ramp in 
Fig.~\ref{fig:DMRGvBOG}\subref{fig:DMRGvBOGa}, long quench times suppress the effect of these early ``jolt'' excitations compared to these caused by the crossing of the critical region.

%

\begin{figure}[htp!]
\vspace{-0cm}
\subfloat{
\renewcommand{\thesubfigure}{a,b}
\includegraphics[width=1.0\columnwidth]{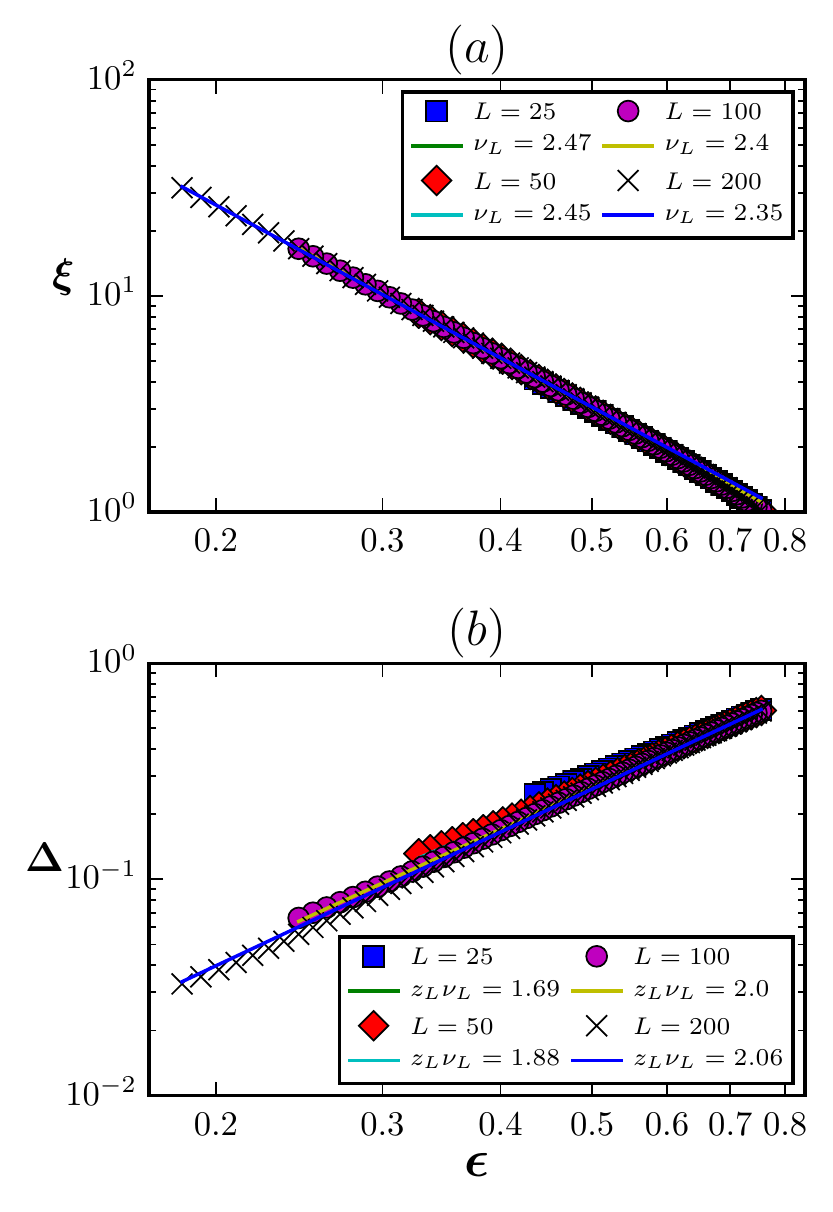}
\label{fig:zLnuLab}
}
\newline
\subfloat{
\includegraphics[width=1.0\columnwidth]{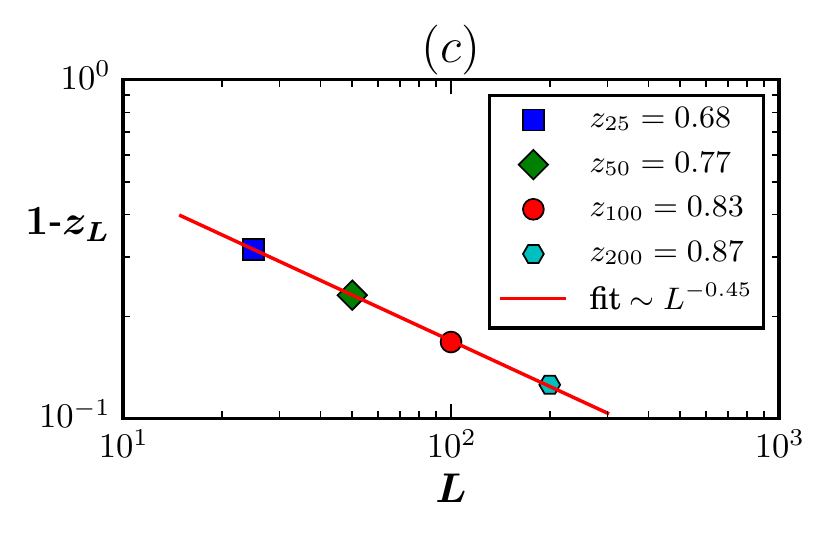}
\renewcommand{\thesubfigure}{c}
\label{fig:zLnuLc}
}
\vspace{-0cm}
\caption{ 
(a) Log-log plot of the correlation length $\xi$ in function of $\epsilon$ for system sizes $L=25,50,100,200$.
For each $L$, the data are fitted with a power law $\xi\sim\epsilon^{-\nu_L}$ in the range of $\epsilon$ where
$1<\xi<L/6$. The best fits are shown by the solid lines and the best exponents $\nu_L$ are listed in the legend. 
In (b), a log-log plot of the energy gap $\Delta$ in function of $\epsilon$ for $L=25,50,100,200$. For each $L$, 
the data are fitted with a power law $\Delta\sim\epsilon^{z_L\nu_L}$ in the range of $\epsilon$ where $1<\xi_L<L/6$. Here $\nu_L$ are the exponents obtained in panel a. The best fits are shown by the solid lines and 
the best exponents $z_L\nu_L$ are listed in the legend. In (c), a log-log plot of $1-z_L$ in function of $L$. 
This plot shows the convergence of $z_L\to1$ with increasing $L$. It can be fitted with $1-z_L\sim L^{-0.45}$ suggesting
a power law approach of $z_L$ towards $z=1$ with increasing $L$.
}
\label{fig:zLnuL}
\end{figure}

The quench times in our numerical simulations are far below the ``astronomical'' standards that would demand $\hat\xi$ of the order of kilometers to approach the scaling $\hat\xi\sim\tau_Q$ in Eq. (\ref{hatxinaive}). Limited by the system size $L$ they span a narrow range of magnitude. Therefore, we can assume a phenomenological power-law for the ground-state correlation length in the Mott phase:
\be 
\xi \sim \epsilon^{-\nu_L}.
\label{nuL}
\ee
Given the exponential divergence~(\ref{xiKT}), we expect the effective exponent $\nu_L$ to be large. To account for finite-size effects, we allow for its dependence on $L$. Furthermore, we found that the data can be accurately parametrized by letting the gap 
scale with an effective dynamical exponent:
\be
\Delta \sim \xi^{-z_L} \sim \epsilon^{z_L\nu_L}.
\label{zL}
\ee
We expect to recover the exact $z_\infty=1$ for sufficiently large $L$.

\begin{figure}[htp!]
\vspace{-1cm}
\includegraphics[width=1.0\columnwidth]{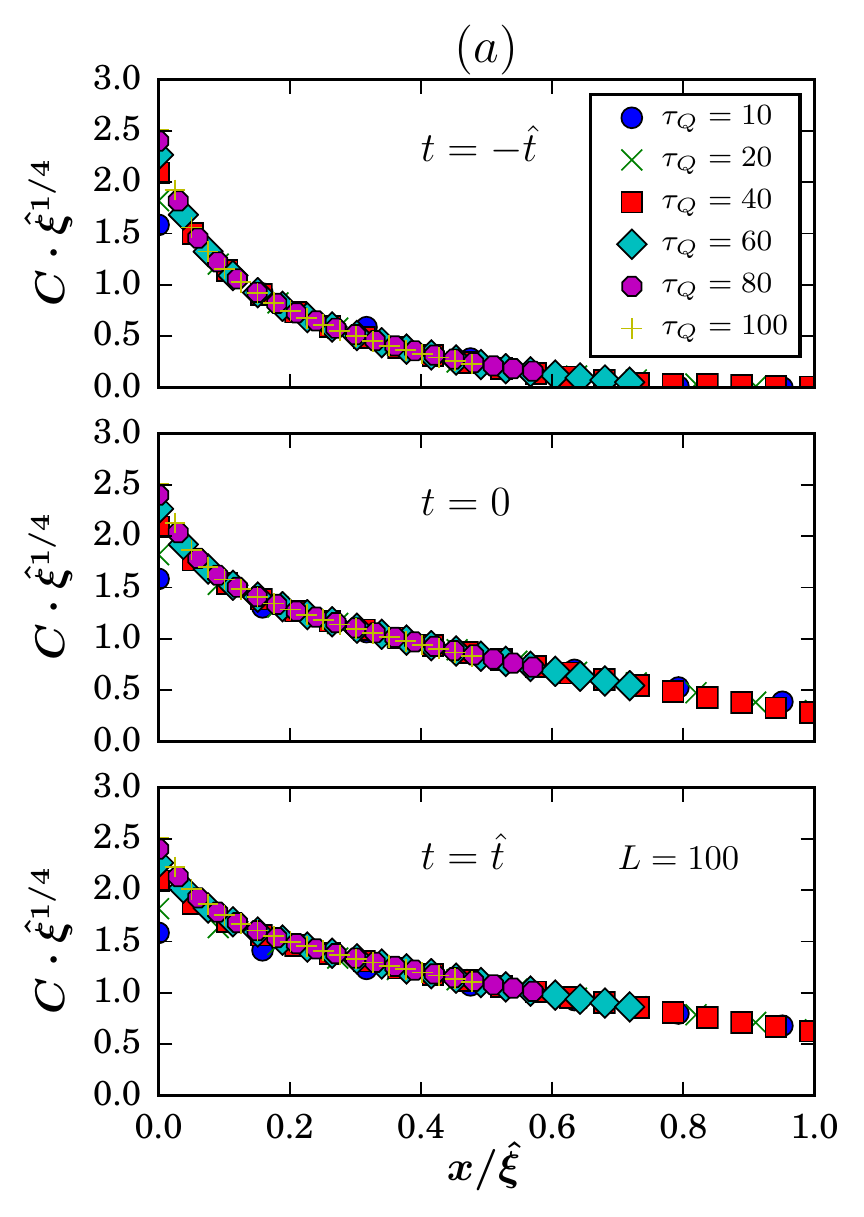}
\includegraphics[width=1.0\columnwidth]{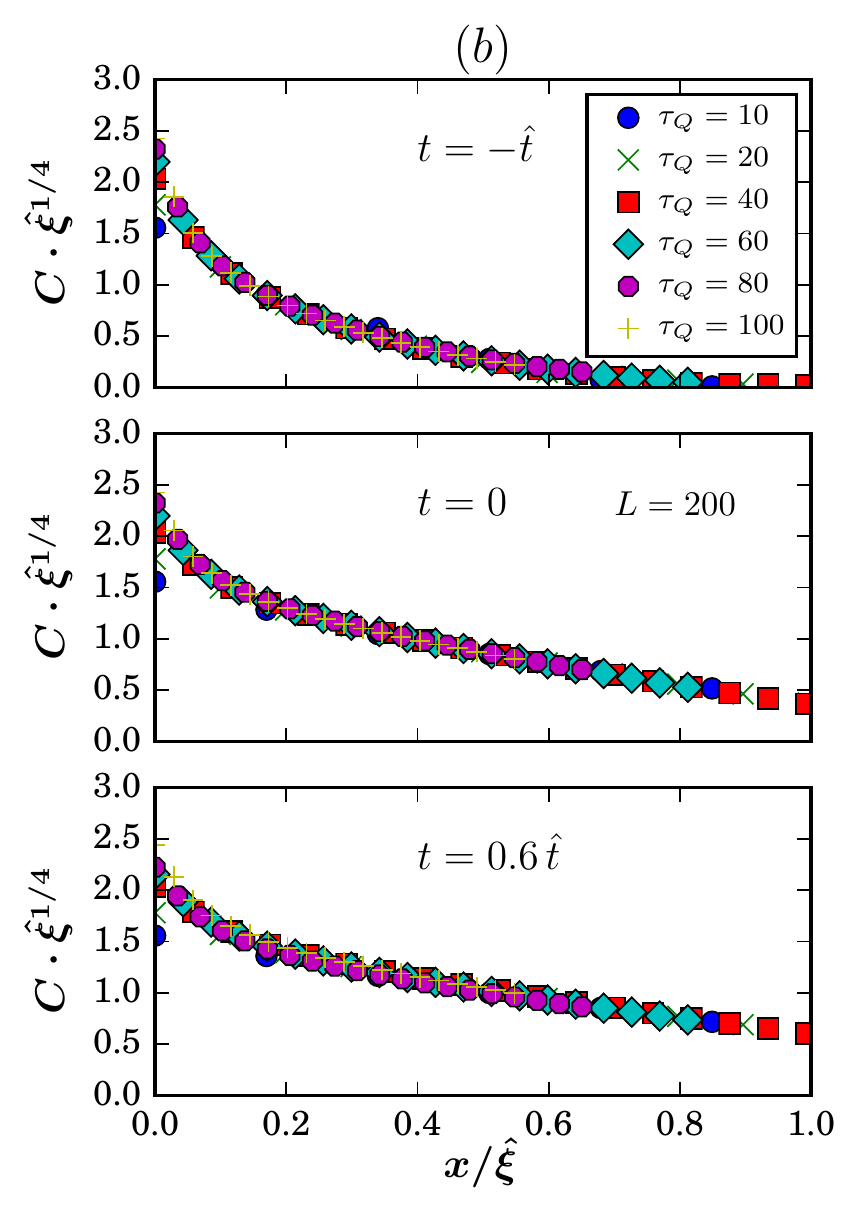}
\vspace{-1cm}
\caption{ 
(a) Linear plots of the scaled correlation functions $\hat\xi^{1/4}C(t,x)$ in function of the scaled distance $x/\hat\xi$ -- measured in the bulk of $L=100$ lattice sites -- at the scaled times $t/\hat t=-1,0,1$. With increasing $\tau_Q
$ the plots collapse to the scaling function $F_C(t/\hat t)$. In (b), the same as in (a) but for $L=200$ and $t/\hat t=-1,0,0.6$. 
}
\label{fig:MISFscaling}
\end{figure}

The correlation length in the ground state is obtained from the best fit to the tail of the correlator:
\be 
C(x) = \langle b_s^\dag b_{s+x} \rangle \sim \frac{e^{-x/\xi}}{x^{1/4}}.
\ee
The lengths and gaps for different $\epsilon$ and $L$ are collected in Fig.~\ref{fig:zLnuL}\subref{fig:zLnuLab}. 
They fit well the phenomenological power-laws~(\ref{nuL},\ref{zL}). As a self-consistency check,
figure~\ref{fig:zLnuL}\subref{fig:zLnuLc} shows how the dynamical exponent $z_L$ decays to $1$ with increasing $L$. For each $L$ we use the effective exponents to find the KZ scales
\be 
\label{hattxiL}
\hat t  \sim \tau_Q^{\frac{z_L\nu_L}{1+z_L\nu_L}}, \quad
\hat\xi \sim \tau_Q^{\frac{\nu_L}{1+z_L\nu_L}}. 
\ee
These scales are applied to verify the KZ scaling hypothesis, 
\be
C(t,x) = \hat\xi^{-1/4} F_C\left(t/\hat\xi^{z_L},x/\hat\xi\right),
\ee
in Fig. \ref{fig:MISFscaling}. The plots for different $\tau_Q$ collapse demonstrating validity of the hypothesis.

\begin{figure}[htp!]
\vspace{-0cm}
\subfloat{
\includegraphics[width=1.0\columnwidth,clip=true]{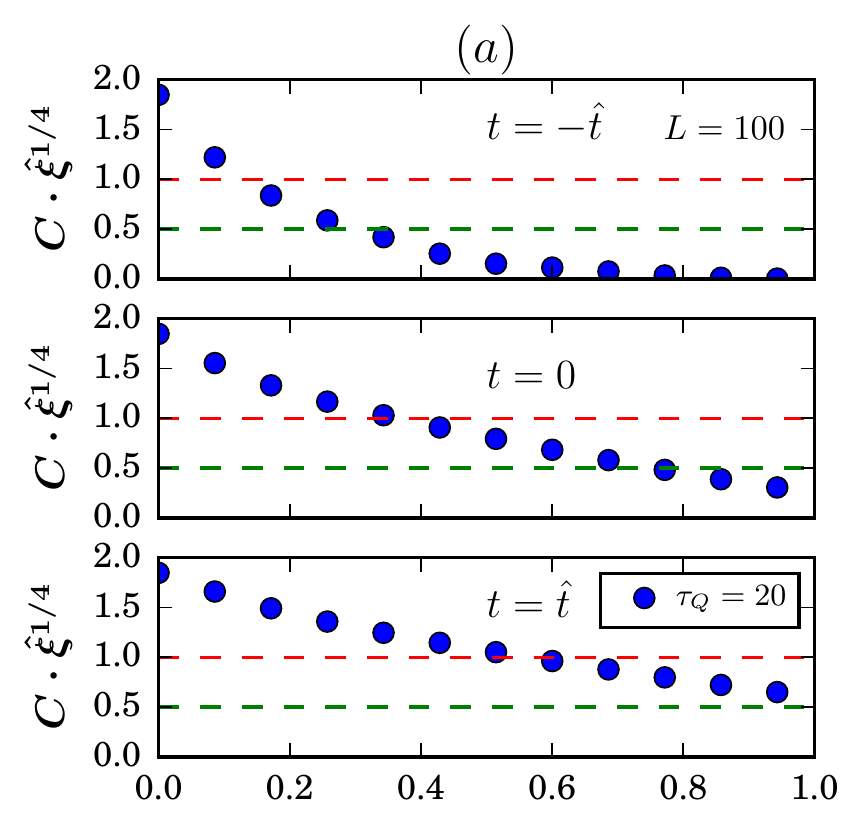}
\label{fig:MISFhatva}
}
\newline
\subfloat{
\includegraphics[width=1.0\columnwidth,clip=true]{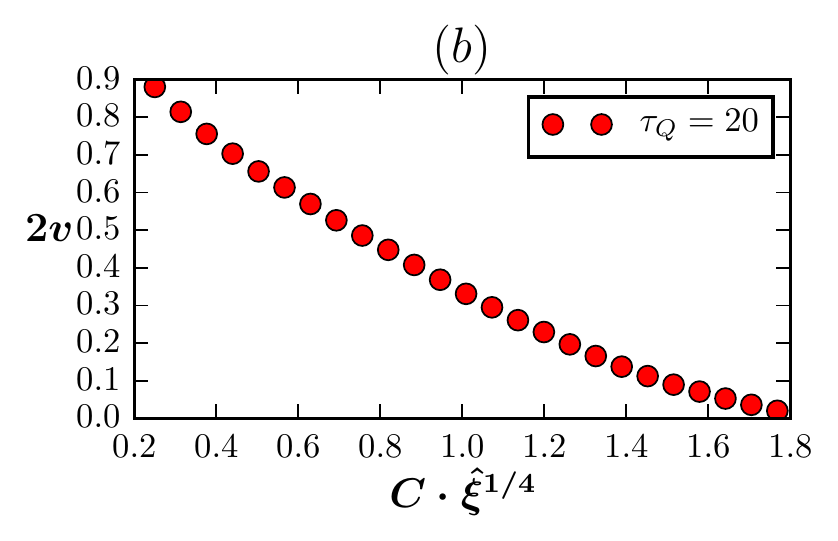}
\label{fig:MISFhatvb}
}
\newline
\subfloat{
\includegraphics[width=1.0\columnwidth,clip=true]{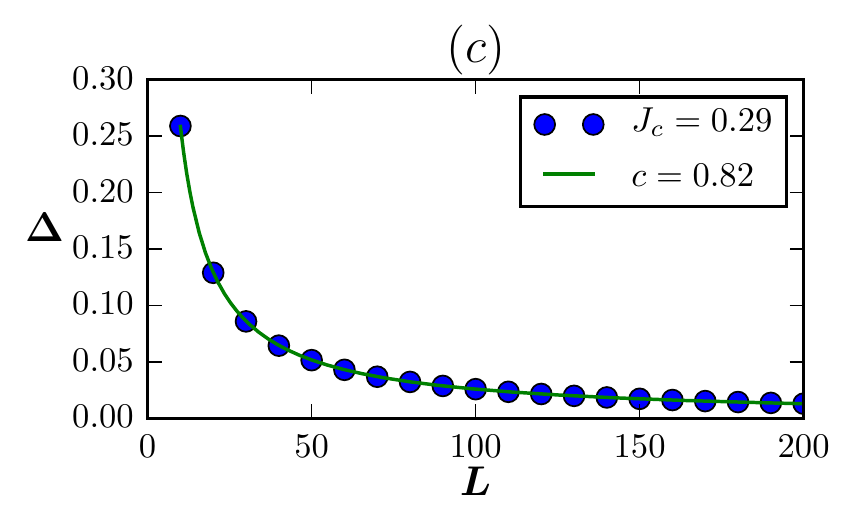}
\label{fig:MISFhatvc}
}
\vspace{-0cm}
\caption{ 
(a) Scaled correlation functions for $\tau_Q=20$ in function of the scaled distance $x/\hat\xi$ measured in the bulk 
of $L=100$ lattice sites. The three plots show the correlator at scaled times: $t/\hat t=-1,0,1$. A cut
by the horizontal dashed line at the level $1$ suggests that the correlations are spreading and the sonic horizon grows 
at a velocity $2\hat v=0.34$, but a similar cut at $0.5$ suggests $2\hat v=0.63$. Longer tails of the correlator, that could be 
probed at lower cut levels, are not available on $L=100$ sites. In (b), spreading velocity $2\hat v$ in function of scaled
correlation values (the cut level). In (c), the gap $\Delta$ at $J_c$ in function of system size $L$. The best fit 
$\Delta=c_0\frac{\pi}{L}$ yields $2c_0=1.64$ as the speed of Luttinger quasiparticles. As expected, $2c_0$ is 
greater than our crude estimates of $2\hat v$.
}
\label{fig:MISFhatv}
\end{figure}

The collapsed plots in Fig.~\ref{fig:MISFscaling} depend on the scaled time $t/\hat t$. This effect reveals that the impulse approximation is not quite correct: since the correlations are spreading, the state cannot be frozen. This behavior suggests a ``sonic horizon'' paradigm we have also noted earlier. Nevertheless, as discussed and demonstrated by an exact solution of the quantum Ising chain~\cite{Francuzetal}, the scaling hypothesis still holds because the velocity $2\hat v$ of the spreading must also be a combination of the two KZ scales: $\hat v\sim\hat\xi/\hat t$. For a linear dispersion at the critical point, $z=1$, $\hat v$ is expected to be bounded from above by twice the speed $c_0$ of quasiparticles at the critical point. In figure~\ref{fig:MISFhatv}\subref{fig:MISFhatva}, we attempt an estimate of the spreading velocity $2\hat v$. Since accurate tails of the correlators cannot be accessed on a finite lattice, our estimate is not robust but at least it is less than the $2c_0$ estimated from the quasiparticle dispersion in figure~\ref{fig:MISFhatv}\subref{fig:MISFhatvb}.

\begin{figure}[htp!]
\vspace{-0cm}
\includegraphics[width=1.0\columnwidth,clip=true]{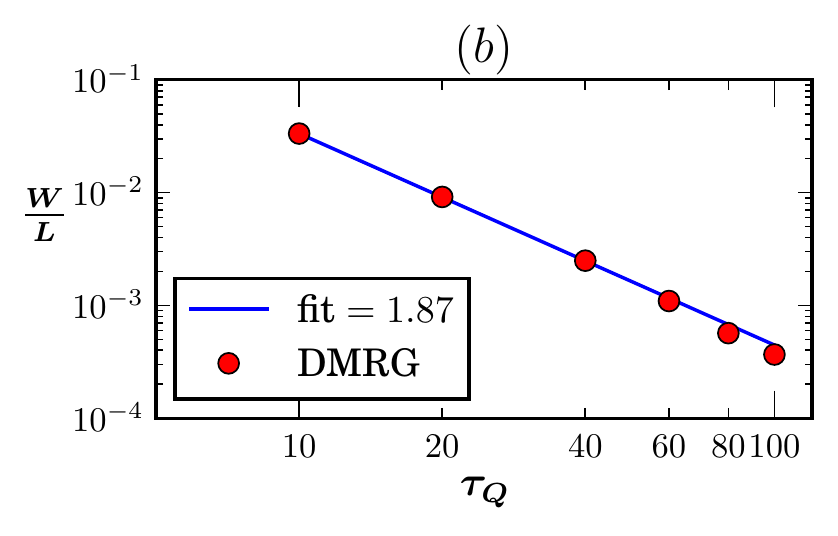}
\includegraphics[width=1.0\columnwidth,clip=true]{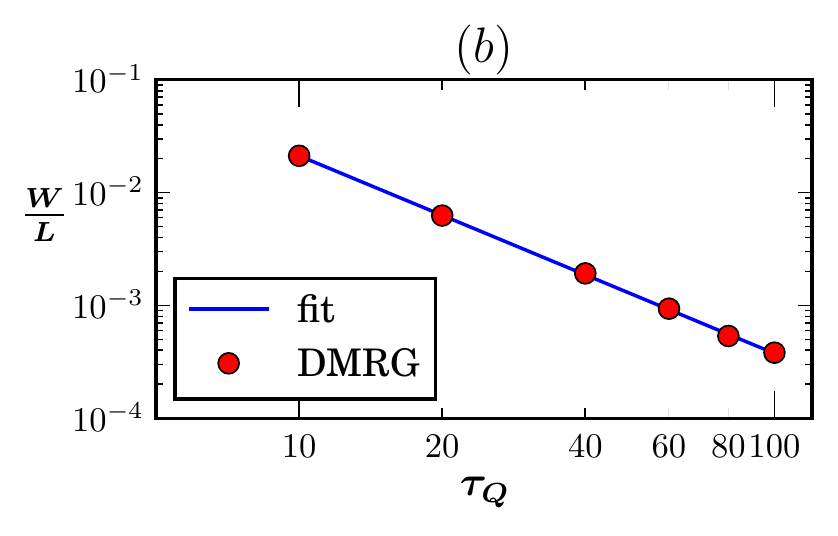}
\includegraphics[width=1.0\columnwidth,clip=true]{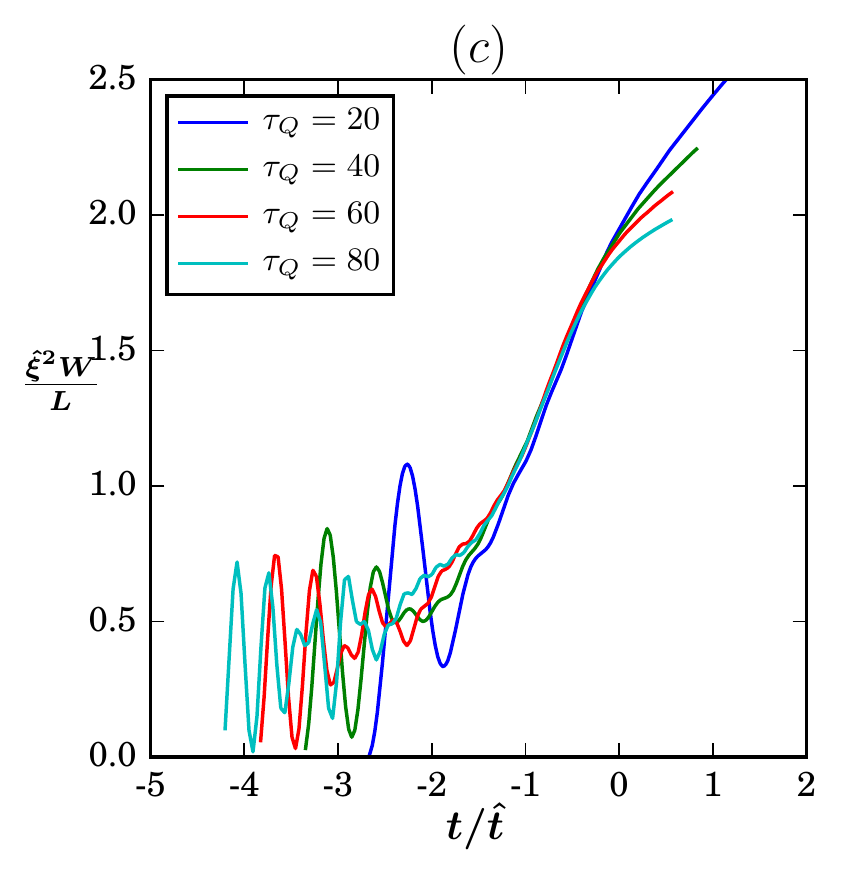}
\vspace{-0cm}
\caption{ 
(a) Log-log plot of the excitation energy density $W/L$ in function of the quench time $\tau_Q$ at the center of $L=100$ lattice at the time $t=0$ when the quench is crossing the critical point. The solid line is the best fit $W\sim\tau_Q^{-1.79}$ consistent with $W\sim\tau_Q^{-1.80}$ predicted by Eqs.~(\ref{hattxiL},\ref{Wtscaling}) with the exponents fitted in Fig.~\ref{fig:zLnuL}. 
(b) The same as in (a) but for $L=200$ sites. The solid line is the best fit $W\sim\tau_Q^{-1.75}$ consistent with 
$W\sim\tau_Q^{-1.77}$ predicted by Eqs.~(\ref{hattxiL},\ref{Wtscaling}) with the exponents fitted in Fig.~\ref{fig:zLnuL}b.
(c) Scaled excitation energy density in function of scaled time for $L=100$ sites. The plots collapse to the scaling function $F_W$ in the KZ regime close to $J_c$. 
}
\label{fig:MISFenergy}
\end{figure}

Neglecting remnants of the early excitation, a distribution of quasiparticle excitations should have a scaling form 
$f(t/\hat t,\hat\xi k)$. In the KZ regime close to $J_c$, their dispersion can be approximated by the gapless linear dispersion at the critical pojnt, $\omega_k\approx c_0 |k|$, and the excitation energy (also sometimes referred to as work~\cite{Binder15,Gardas15}) density should satisfy a scaling hypothesis:
\be 
W/L = \int \frac{dk}{2\pi}~ c_0|k|~ f(t/\hat t,\hat\xi k) = \hat\xi^{-2} F_W\left(t/\hat t\right).
\label{Wtscaling}
\ee
Here $F_W$ is a scaling function. In particular, at $J_c$ (where $t/\hat t=0$) we expect $W\sim\hat\xi^{-2}$. This scaling law and the more general scaling hypothesis (\ref{Wtscaling}) are confirmed by the data in Fig. \ref{fig:MISFenergy}. 

\begin{figure}[htp!]
\vspace{-0cm}
\includegraphics[width=1.0\columnwidth,clip=true]{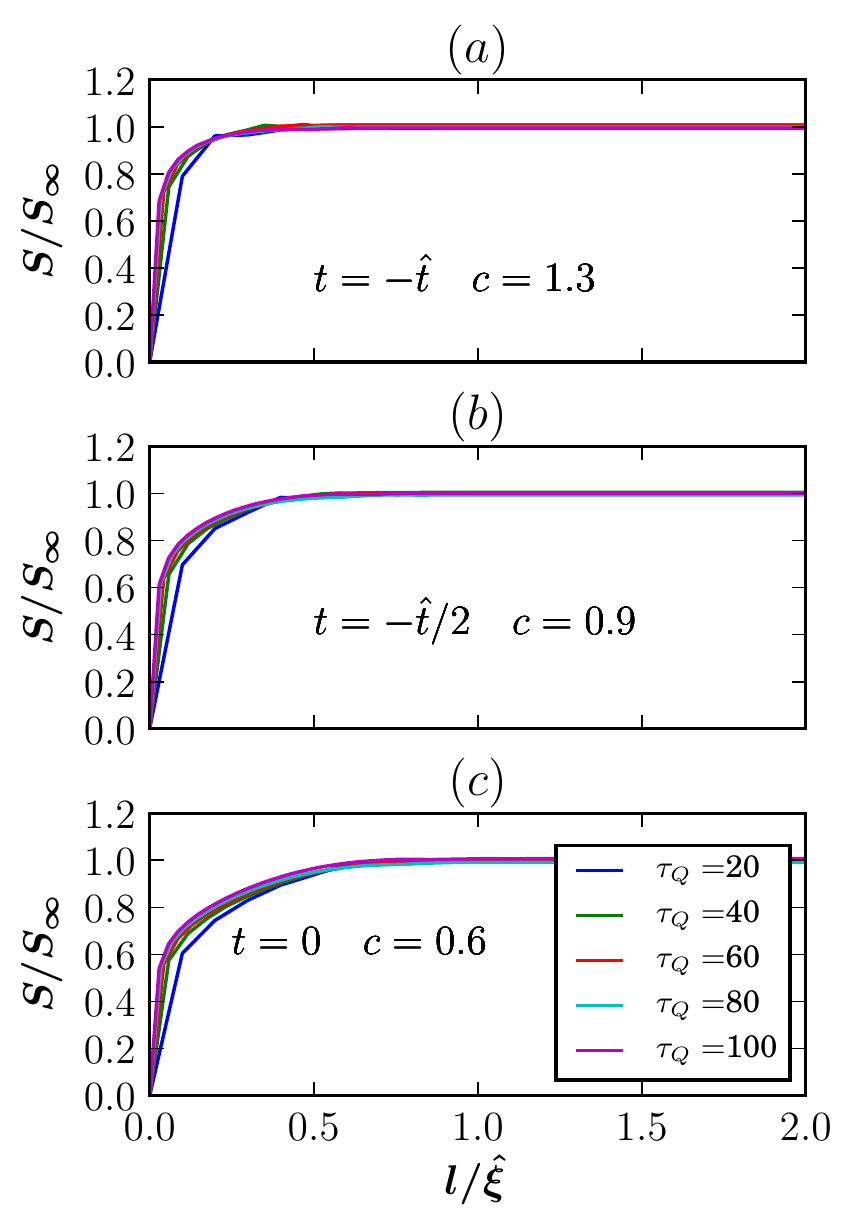}
\vspace{-0cm}
\caption{ 
Entanglement entropy $S(t,l)$ between a block of $l$ sites at the end and the rest of the chain of $L=200$ sites. 
Panel (a), (b), and (c) show the entropy at $t=-\hat t$, $-\hat t/2$, and $0$, respectively. All plots are in function of the scaled block size, $l/\hat\xi$, and the entropy is divided by 
$
S_\infty(t/\hat t)=
\frac{c}{6}\ln \hat\xi f.
$
This logarithmic function of $\tau_Q$ is obtained as the best fit to $S\left(t/\hat t,L/2\right)$ with fitting parameters $f$ and $c$.
The best central charge $c$ for $t=-\hat t$, $-\hat t/2$, $0$ is $c=1.3$, $0.9$, $0.6$, respectively.
The plots collapse demonstrating the scaling hypothesis in Eq.~(\ref{Sscaling}).
}
\label{fig:MISFentropy}
\end{figure}


Finally, we considered the entropy of entanglement between a block of $l$ sites at the end of the chain and the rest of it. Ignoring boundary effects, the entropy should satisfy a scaling hypothesis: 
\be 
\frac{ S(t,l) }{ \frac{c}{6} \ln \hat\xi f(t/\hat t) } =
F_S\left(t/\hat t,l/\hat\xi\right).
\label{Sscaling}
\ee
Here $f$ and $F_S$ are scaling functions and $c$ is the central charge $c=1$ at the Kosterlitz-Thouless transition. This hypothesis is tested in Fig. \ref{fig:MISFentropy}. Except for $t=-\hat t$, where the state is still close to the ground state, the fitted $c$, though close to $1$, are not quite satisfactory, but we have to bear in mind that the range of $\tau_Q$ at hand is too narrow for a better fit with a logarithmic function. However, we find that -- in accordance with the scaling hypothesis -- the plots collapse to a scaling function $F_S$ after rescaling the block size by $\hat\xi$.
 
\section{Quench from superfluid to Mott insulator}
\label{sec:SFMI}

In this section we reverse the quench. Now the linear ramp begins when $t=-\tau_Q$ deep in the superfluid phase at $J=2J_c$ and ends at $J=0$ when $t=\tau_Q$:
\be 
J(t)=J_c\left(1-t/\tau_Q\right).
\label{eq:Jt2}
\ee
The initial ground state has a quasi-long-range order characterized by a power-law decay of its correlation function 
$C^{(GS)}(x)$. Small excitations in the superfluid -- described by a Luttinger liquid -- are gapless, hence it may be tempting to treat the whole evolution in the superfluid phase as impulse. However, different critical ground states in the superfluid phase are sufficiently similar to each other for a significant excitation to be postponed until after the ramp crosses the boundary with the Mott phase where the gap begins to open and the ground state begins to change fundamentally. From this perspective, the similarity between different superfluid critical states means that on the superfluid side of the transition there is little difference between the impulse and adiabatic approximations. The whole superfluid critical phase can be effectively collapsed to the Mott critical point.

\begin{figure}[htp!]
\vspace{-0cm}
\subfloat{
\includegraphics[width=1.0\columnwidth,clip=true]{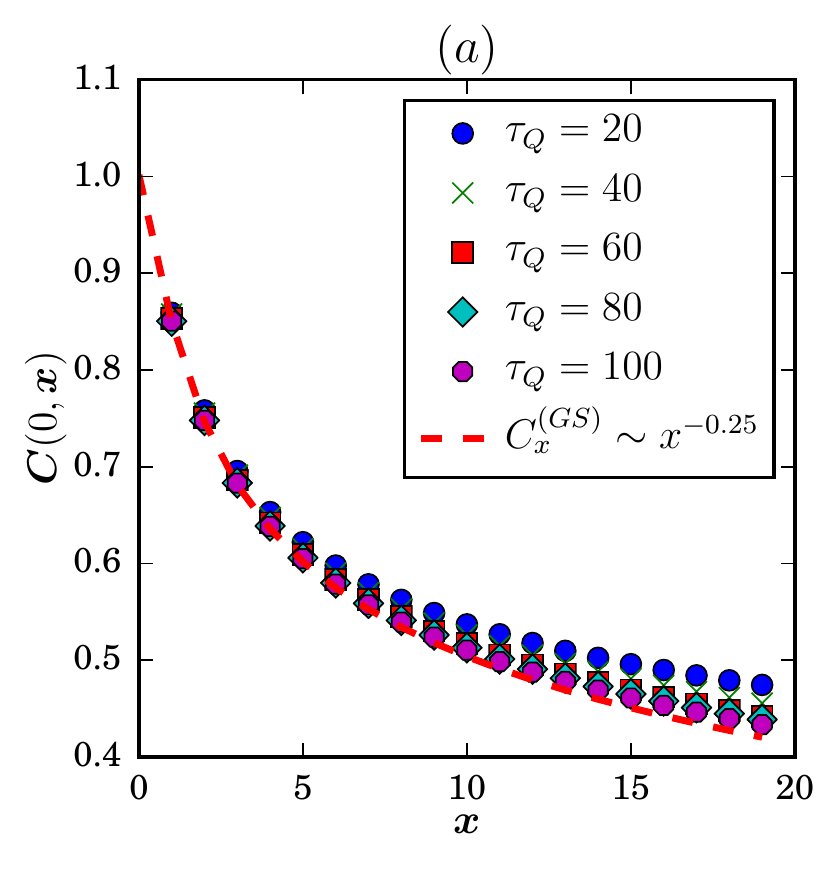}
\label{fig:SFMICxt0a}
}
\newline
\subfloat{
\includegraphics[width=1.0\columnwidth,clip=true]{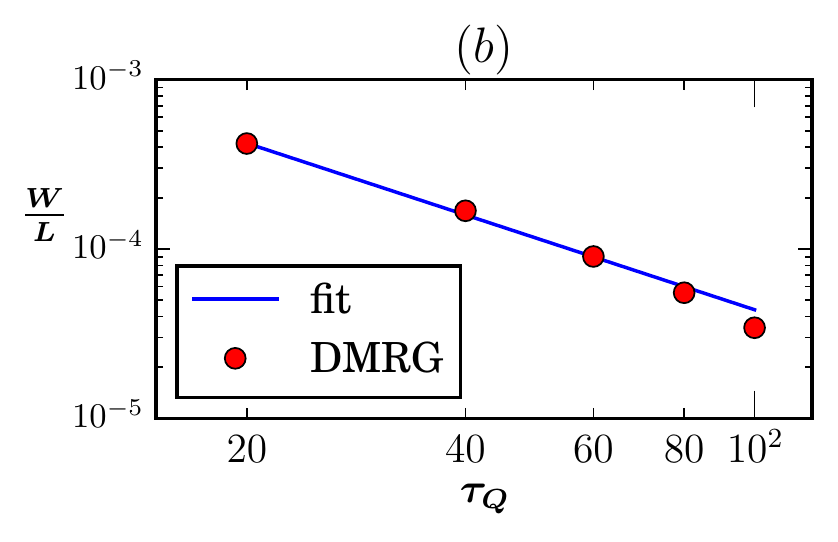}
\label{fig:SFMICxt0b}
}
\vspace{-0cm}
\caption{ 
(a) Correlation functions $C(0,x)$ at the critical point, $J=J_c$, during a linear quench (\ref{eq:Jt2}) from the superfluid at $J=2J_c$ to the Mott insulator at $J=0$. With increasing $\tau_Q$ they converge to the adiabatic ground state correlator with a power-law tail: $C_x^{(GS)}\sim x^{-1/4}$. (b) Log-log plot of the corresponding excitation energy density $W/L$ at the center of $L=100$ lattice sites. These data are fitted with a power law $W\sim\tau_Q^{-1.41}$.
}
\label{fig:SFMICxt0}
\end{figure}

\begin{figure}[htp!]
\vspace{-0cm}
\subfloat{
\includegraphics[width=1.0\columnwidth,clip=true]{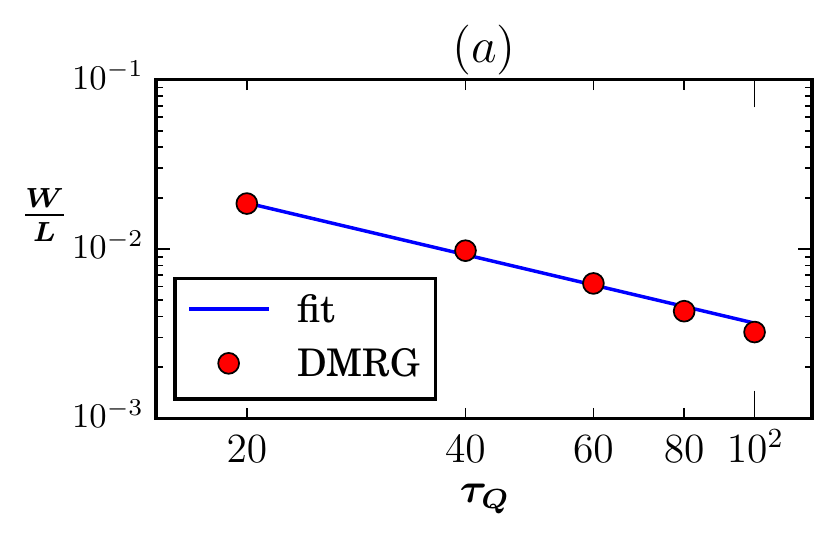}
\label{fig:SFMIWJ0a}
}
\newline
\subfloat{
\includegraphics[width=1.0\columnwidth,clip=true]{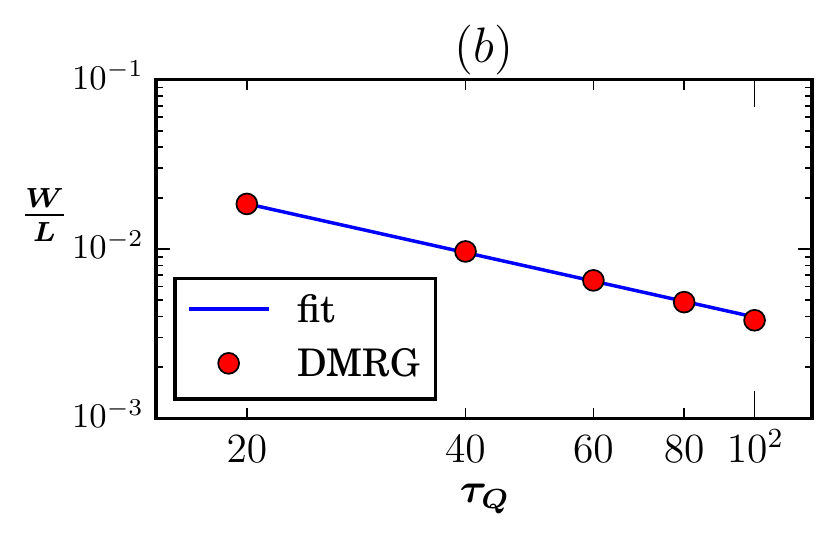}
\label{fig:SFMIWJ0b}
}
\vspace{-0cm}
\caption{ 
(a) Excitation energy density $W/L$ in function of the quench time $\tau_Q$ at $J=0$ after the full linear ramp, $J(t)=J_c\left(1-t/\tau_Q\right)$, starting deep in the superfluid phase at $J=2J_c$. (b) The same as in (a) but after a half-ramp starting from the critical point $J=J_c$. The data in (a) and (b) are almost the same as if the initial half of the full ramp in the superfluid phase did not contribute to the final excitation energy deep in the Mott phase.
}
\label{fig:SFMIWJ0}
\end{figure}

Indeed, at $t=0$ -- when the quench is leaving the superfluid at $J_c$ -- the correlator $C(t,x)$ quickly converges with increasing $\tau_Q$ to the correlator in the adiabatic ground state $C^{(GS)}(x)\sim x^{-1/4}$ at $J_c$, see figure~\ref{fig:SFMICxt0}\subref{fig:SFMICxt0a}. This fast convergence is consistent with the quick decay of the excitation energy density shown in Fig.~\ref{fig:SFMICxt0}\subref{fig:SFMICxt0b}. Its decay is steep, almost as steep as predicted for a linear ramp in a Luttinger liquid \cite{TylutkiLutt}: $W\sim\tau_Q^{-2}\ln(c_0\tau_Q)$ with $c_0\simeq 1$ standing for the speed of Luttinger quasiparticles. It turns out to be steep enough for the excitation in the superfluid phase to give negligible contribution to the final excitation deep in the Mott phase, see Fig.~\ref{fig:SFMIWJ0}. Its panel~\subref{fig:SFMICxt0a} shows the final excitation energy  density after the full ramp from $J=2J_c$ to $J=0$ and panel~\subref{fig:SFMICxt0b} after a shorter ramp from $J=J_c$ to $J=0$. The two panels are almost indistinguishable and in both the excitation energy density is two orders of magnitude higher than in Fig. \ref{fig:SFMICxt0}\subref{fig:SFMICxt0b}. Therefore, the final excitation after the full ramp originates almost exclusively from the evolution in the Mott phase when the gap opens and the ground state undergoes a fundamental change. 

%
\begin{figure}[htp!]
	\begin{tikzpicture}[scale=1.0] 
	\draw [ultra thick, ->] (0,0)--(8,0);	
	\draw [ultra thick, ->] (0,0)--(0,4.);
	\node [right] at (8,0) {\large $t$};
	\draw [thick] (0,-.1)  node[below]{\large $0$}  	  -- (0,0.1);
	\draw [thick] (3.0,-.1)  node[below]{\large $\hat{t}$} -- (3.0,0.1);
	\node at (+6,4.3) {adiabatic};
	\node at (1.4,4.3) {impulse};
	\node[scale=1.6] at (+1.2,3.5)    {\small \bl{rate}};
	\node[scale=1.6] at (+6.0,0.8) {\small \gn{gap}};
	\draw[scale=0.6,domain=0.6:12,smooth,variable=\x,blue] plot ({\x},{+4./\x});
	\draw[scale=0.6,domain=0.001:12,smooth,variable=\x,green] plot ({\x},{exp(-0.5/sqrt(\x))});	
	\draw[ultra thick,dashed,red]  (3,0) -- (3,4);
	\fill[gray!20,nearly transparent] (0,0) -- (3,0) -- (3,4) -- (0,4) -- cycle;
	\end{tikzpicture}
	\caption{
	A ramp from the critical point at $J=J_c$ to $J=0$ deep in the Mott phase. 
	The evolution crosses over from impulse to adiabatic near $\hat t$.
	}	
	\label{fig:KZMhalf}
\end{figure}
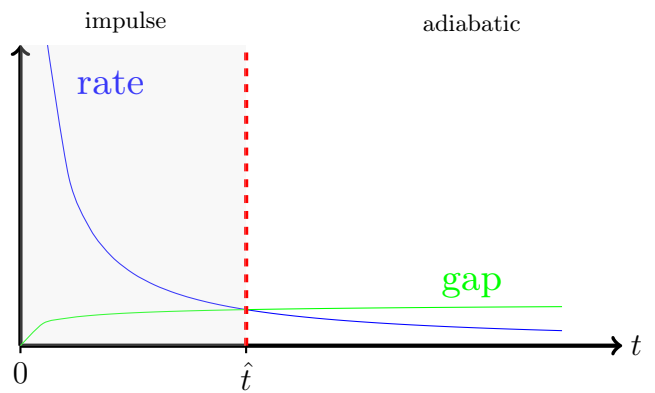

The Mott part of the full ramp, from $J_c$ to $0$, falls within the KZ framework, as shown schematically in figure \ref{fig:KZMhalf}. The evolution in the Mott phase crosses over from impulse to adiabatic near $\hat t\sim\tau_Q^{\frac{z_L\nu_L}{1+z_L\nu_L}}$ when the correlation length in the ground state is $\hat\xi\sim\tau_Q^{\frac{\nu_L}{1+z_L\nu_L}}$. As long as quasiparticle excitations can be considered non-interacting, their distribution $f(t/\hat t,\hat\xi k)$ satisfies the KZ scaling hypothesis. This scaling form has at least two important consequences.

\begin{figure}[htp!]
	\vspace{-0cm}
	\includegraphics[width=1.0\columnwidth,clip=true]{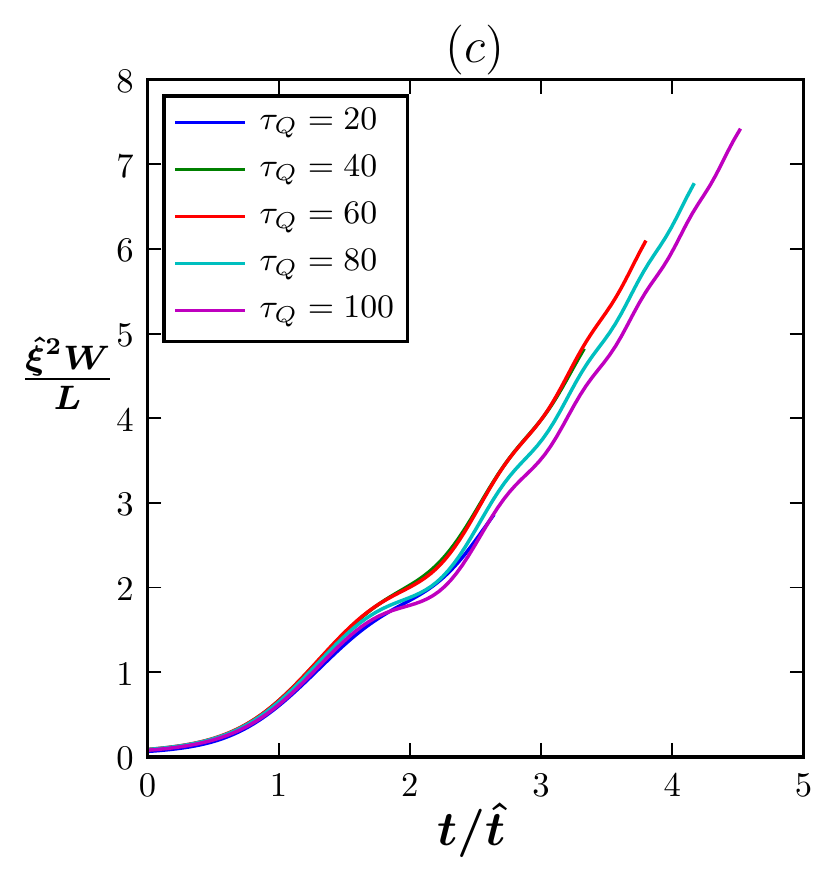}
	\vspace{-0cm}
	\caption{ 
		Scaled excitation energy  density $\hat\xi^2 W/L$ in function of scaled time $t/\hat t$ for different quench times $\tau_Q$. The plots collapse to the scaling function $F_W(t/\hat t)$ in Eq. (\ref{Wtscaling}).
	}
	\label{fig:SFMIwt}
\end{figure}

Before $\hat t$, when the quasiparticle dispersion can be approximated by its critical form $\omega_k=c_0k$, the excitation energy  density should conform to the scaling law (\ref{Wtscaling}). Indeed, the scaled plots in Fig. \ref{fig:SFMIwt} demonstrate a collapse to the scaling function $F_W$. As predicted, the collapse is perfect up to $\hat t$.

In the adiabatic stage after $\hat t$, the quasiparticle distribution freezes out, 
$f(t/\hat t,\hat\xi k)=f(\hat\xi k)$, and their dispersion $\omega_k$ for the ``excited'' $k$ between $\pm\hat\xi^{-1}$ can be approximated by the finite gap $\omega_0$. Consequently, the final excitation energy  density at $J=0$ should scale as
\be 
\frac{W}{L} = 
\int_{-\pi}^\pi \frac{dk}{2\pi} ~ \omega_k ~ f\left(\hat\xi k\right) \sim 
\hat\xi^{-1}.
\label{WJscaling}
\ee
Since in this regime the excited quasiparticles are approximately dispersionless,
$\omega_k\approx\omega_0$, the excitation energy density is simply proportional to their density $\hat\xi^{-1}$. With our best fits for $\nu_L$ and $z_L$ on the $L=100$ lattice we obtain $W\sim\tau_Q^{-0.80}$. This is roughly consistent with the best fit $W\sim\tau_Q^{-0.96}$ in Fig.~\ref{fig:SFMIWJ0}\subref{fig:SFMIWJ0b}. 

\section{Conclusion}
\label{sec:concl}
In a linear quench from Mott insulator to superfluid the excitation energy  density, the entropy of entanglement, and the correlations -- that build up as the system is crossing the critical point -- satisfy the KZ scaling hypothesis with effective power laws accurate for a limited range of quench times \cite{KZSR}. In particular, the range of correlations scales with an effective power of the transition time.

All superfluid ground states are qualitatively similar. Therefore, in a reverse quench from superfluid to Mott insulator the excitation in the gapless superfluid turns out to be negligible as compared to the excitation that begins to build up just after crossing the critical point when the gap opens and the ground state begins to change fundamentally. The last excitation also falls into the KZ framework. The final excitation energy deep in the Mott phase -- proportional to the number of empty and 
doubly-occupied sites -- decays with an effective power of the quench time.
%
\acknowledgments
%
Work of J.D. and B.G. was supported in part by Narodowe Centrum Nauki (National Science 
Center) under Project No. 2013/09/B/ST3/01603 and 2016/20/S/ST2/00152, respectively. 
Work of W.H.Z. was supported by the US Department of Energy under the Los Alamos LDRD program.
This research was supported in part by PL-Grid Infrastructure.
%
%

\begin{thebibliography}{101}%
	\makeatletter
	\providecommand \@ifxundefined [1]{%
		\@ifx{#1\undefined}
	}%
	\providecommand \@ifnum [1]{%
		\ifnum #1\expandafter \@firstoftwo
		\else \expandafter \@secondoftwo
		\fi
	}%
	\providecommand \@ifx [1]{%
		\ifx #1\expandafter \@firstoftwo
		\else \expandafter \@secondoftwo
		\fi
	}%
	\providecommand \natexlab [1]{#1}%
	\providecommand \enquote  [1]{``#1''}%
	\providecommand \bibnamefont  [1]{#1}%
	\providecommand \bibfnamefont [1]{#1}%
	\providecommand \citenamefont [1]{#1}%
	\providecommand \href@noop [0]{\@secondoftwo}%
	\providecommand \href [0]{\begingroup \@sanitize@url \@href}%
	\providecommand \@href[1]{\@@startlink{#1}\@@href}%
	\providecommand \@@href[1]{\endgroup#1\@@endlink}%
	\providecommand \@sanitize@url [0]{\catcode `\\12\catcode `\$12\catcode
		`\&12\catcode `\#12\catcode `\^12\catcode `\_12\catcode `\%12\relax}%
	\providecommand \@@startlink[1]{}%
	\providecommand \@@endlink[0]{}%
	\providecommand \url  [0]{\begingroup\@sanitize@url \@url }%
	\providecommand \@url [1]{\endgroup\@href {#1}{\urlprefix }}%
	\providecommand \urlprefix  [0]{URL }%
	\providecommand \Eprint [0]{\href }%
	\providecommand \doibase [0]{http://dx.doi.org/}%
	\providecommand \selectlanguage [0]{\@gobble}%
	\providecommand \bibinfo  [0]{\@secondoftwo}%
	\providecommand \bibfield  [0]{\@secondoftwo}%
	\providecommand \translation [1]{[#1]}%
	\providecommand \BibitemOpen [0]{}%
	\providecommand \bibitemStop [0]{}%
	\providecommand \bibitemNoStop [0]{.\EOS\space}%
	\providecommand \EOS [0]{\spacefactor3000\relax}%
	\providecommand \BibitemShut  [1]{\csname bibitem#1\endcsname}%
	\let\auto@bib@innerbib\@empty
	\bibitem [{\citenamefont {Kibble}(1976)}]{Kibble76}%
	\BibitemOpen
	\bibfield  {author} {\bibinfo {author} {\bibfnamefont {T.~W.~B.}\
			\bibnamefont {Kibble}},\ }\href {http://stacks.iop.org/0305-4470/9/i=8/a=029}
	{\bibfield  {journal} {\bibinfo  {journal} {J. Phys. A: Math. Gen}\ }\textbf
		{\bibinfo {volume} {9}},\ \bibinfo {pages} {1387} (\bibinfo {year}
		{1976})}\BibitemShut {NoStop}%
	\bibitem [{\citenamefont {Kibble}(1980)}]{Kibble80}%
	\BibitemOpen
	\bibfield  {author} {\bibinfo {author} {\bibfnamefont {T.~W.~B.}\
			\bibnamefont {Kibble}},\ }\href {\doibase
		http://dx.doi.org/10.1016/0370-1573(80)90091-5} {\bibfield  {journal}
		{\bibinfo  {journal} {Phys. Rep.}\ }\textbf {\bibinfo {volume} {67}},\
		\bibinfo {pages} {183} (\bibinfo {year} {1980})}\BibitemShut {NoStop}%
	\bibitem [{\citenamefont {Hindmarsh}(2016)}]{Hindmarsh17}%
	\BibitemOpen
	\bibfield  {author} {\bibinfo {author} {\bibfnamefont {M.}~\bibnamefont
			{Hindmarsh}},\ }\href {http://stacks.iop.org/1751-8121/49/i=41/a=411001}
	{\bibfield  {journal} {\bibinfo  {journal} {J. Phys. A: Math. Theor.}\
		}\textbf {\bibinfo {volume} {49}},\ \bibinfo {pages} {411001} (\bibinfo
		{year} {2016})}\BibitemShut {NoStop}%
	\bibitem [{\citenamefont {Zurek}(1985)}]{Zurek85}%
	\BibitemOpen
	\bibfield  {author} {\bibinfo {author} {\bibfnamefont {W.~H.}\ \bibnamefont
			{Zurek}},\ }\href {\doibase doi:10.1038/317505a0} {\bibfield  {journal}
		{\bibinfo  {journal} {Nature}\ }\textbf {\bibinfo {volume} {317}},\ \bibinfo
		{pages} {505} (\bibinfo {year} {1985})}\BibitemShut {NoStop}%
	\bibitem [{\citenamefont {Zurek}(1993)}]{Zurek93}%
	\BibitemOpen
	\bibfield  {author} {\bibinfo {author} {\bibfnamefont {W.~H.}\ \bibnamefont
			{Zurek}},\ }\href
	{http://www.actaphys.uj.edu.pl/vol24/abs/v24p1301.htm?series=reg&vol=24&page=1301}
	{\bibfield  {journal} {\bibinfo  {journal} {Acta Phys. Pol. B}\ }\textbf
		{\bibinfo {volume} {24}},\ \bibinfo {pages} {1301} (\bibinfo {year}
		{1993})}\BibitemShut {NoStop}%
	\bibitem [{\citenamefont {Zurek}(1996)}]{Zurek96}%
	\BibitemOpen
	\bibfield  {author} {\bibinfo {author} {\bibfnamefont {W.~H.}\ \bibnamefont
			{Zurek}},\ }\href {\doibase doi:10.1016/S0370-1573(96)00009-9} {\bibfield
		{journal} {\bibinfo  {journal} {Phys. Rep.}\ }\textbf {\bibinfo {volume}
			{276}},\ \bibinfo {pages} {177} (\bibinfo {year} {1996})}\BibitemShut
	{NoStop}%
	\bibitem [{\citenamefont {Laguna}\ and\ \citenamefont
		{Zurek}(1997)}]{LagunaZ1}%
	\BibitemOpen
	\bibfield  {author} {\bibinfo {author} {\bibfnamefont {P.}~\bibnamefont
			{Laguna}}\ and\ \bibinfo {author} {\bibfnamefont {W.~H.}\ \bibnamefont
			{Zurek}},\ }\href {\doibase 10.1103/PhysRevLett.78.2519} {\bibfield
		{journal} {\bibinfo  {journal} {Phys. Rev. Lett.}\ }\textbf {\bibinfo
			{volume} {78}},\ \bibinfo {pages} {2519} (\bibinfo {year}
		{1997})}\BibitemShut {NoStop}%
	\bibitem [{\citenamefont {Yates}\ and\ \citenamefont {Zurek}(1998)}]{YZ}%
	\BibitemOpen
	\bibfield  {author} {\bibinfo {author} {\bibfnamefont {A.}~\bibnamefont
			{Yates}}\ and\ \bibinfo {author} {\bibfnamefont {W.~H.}\ \bibnamefont
			{Zurek}},\ }\href {\doibase 10.1103/PhysRevLett.80.5477} {\bibfield
		{journal} {\bibinfo  {journal} {Phys. Rev. Lett.}\ }\textbf {\bibinfo
			{volume} {80}},\ \bibinfo {pages} {5477} (\bibinfo {year}
		{1998})}\BibitemShut {NoStop}%
	\bibitem [{\citenamefont {Dziarmaga}\ \emph {et~al.}(1999)\citenamefont
		{Dziarmaga}, \citenamefont {Laguna},\ and\ \citenamefont {Zurek}}]{DLZ99}%
	\BibitemOpen
	\bibfield  {author} {\bibinfo {author} {\bibfnamefont {J.}~\bibnamefont
			{Dziarmaga}}, \bibinfo {author} {\bibfnamefont {P.}~\bibnamefont {Laguna}}, \
		and\ \bibinfo {author} {\bibfnamefont {W.~H.}\ \bibnamefont {Zurek}},\ }\href
	{\doibase 10.1103/PhysRevLett.82.4749} {\bibfield  {journal} {\bibinfo
			{journal} {Phys. Rev. Lett.}\ }\textbf {\bibinfo {volume} {82}},\ \bibinfo
		{pages} {4749} (\bibinfo {year} {1999})}\BibitemShut {NoStop}%
	\bibitem [{\citenamefont {Antunes}\ \emph {et~al.}(1999)\citenamefont
		{Antunes}, \citenamefont {Bettencourt},\ and\ \citenamefont {Zurek}}]{ABZ99}%
	\BibitemOpen
	\bibfield  {author} {\bibinfo {author} {\bibfnamefont {N.~D.}\ \bibnamefont
			{Antunes}}, \bibinfo {author} {\bibfnamefont {L.~M.~A.}\ \bibnamefont
			{Bettencourt}}, \ and\ \bibinfo {author} {\bibfnamefont {W.~H.}\ \bibnamefont
			{Zurek}},\ }\href {\doibase 10.1103/PhysRevLett.82.2824} {\bibfield
		{journal} {\bibinfo  {journal} {Phys. Rev. Lett.}\ }\textbf {\bibinfo
			{volume} {82}},\ \bibinfo {pages} {2824} (\bibinfo {year}
		{1999})}\BibitemShut {NoStop}%
	\bibitem [{\citenamefont {Bettencourt}\ \emph {et~al.}(2000)\citenamefont
		{Bettencourt}, \citenamefont {Antunes},\ and\ \citenamefont {Zurek}}]{ABZ00}%
	\BibitemOpen
	\bibfield  {author} {\bibinfo {author} {\bibfnamefont {L.~M.~A.}\
			\bibnamefont {Bettencourt}}, \bibinfo {author} {\bibfnamefont {N.~D.}\
			\bibnamefont {Antunes}}, \ and\ \bibinfo {author} {\bibfnamefont {W.~H.}\
			\bibnamefont {Zurek}},\ }\href {\doibase 10.1103/PhysRevD.62.065005}
	{\bibfield  {journal} {\bibinfo  {journal} {Phys. Rev. D}\ }\textbf {\bibinfo
			{volume} {62}},\ \bibinfo {pages} {065005} (\bibinfo {year}
		{2000})}\BibitemShut {NoStop}%
	\bibitem [{\citenamefont {Zurek}\ \emph {et~al.}(2000)\citenamefont {Zurek},
		\citenamefont {Bettencourt}, \citenamefont {Dziarmaga},\ and\ \citenamefont
		{Antunes}}]{BZDA00}%
	\BibitemOpen
	\bibfield  {author} {\bibinfo {author} {\bibfnamefont {W.~H.}\ \bibnamefont
			{Zurek}}, \bibinfo {author} {\bibfnamefont {L.~M.~A.}\ \bibnamefont
			{Bettencourt}}, \bibinfo {author} {\bibfnamefont {J.}~\bibnamefont
			{Dziarmaga}}, \ and\ \bibinfo {author} {\bibfnamefont {N.~D.}\ \bibnamefont
			{Antunes}},\ }\href
	{http://www.actaphys.uj.edu.pl/vol31/abs/v31p2937.htm?series=reg&vol=31&page=2937}
	{\bibfield  {journal} {\bibinfo  {journal} {Acta Phys. Pol. B}\ }\textbf
		{\bibinfo {volume} {31}},\ \bibinfo {pages} {2937} (\bibinfo {year}
		{2000})}\BibitemShut {NoStop}%
	\bibitem [{\citenamefont {del Campo}\ \emph {et~al.}(2010)\citenamefont {del
			Campo}, \citenamefont {De~Chiara}, \citenamefont {Morigi}, \citenamefont
		{Plenio},\ and\ \citenamefont {Retzker}}]{ions20}%
	\BibitemOpen
	\bibfield  {author} {\bibinfo {author} {\bibfnamefont {A.}~\bibnamefont {del
				Campo}}, \bibinfo {author} {\bibfnamefont {G.}~\bibnamefont {De~Chiara}},
		\bibinfo {author} {\bibfnamefont {G.}~\bibnamefont {Morigi}}, \bibinfo
		{author} {\bibfnamefont {M.~B.}\ \bibnamefont {Plenio}}, \ and\ \bibinfo
		{author} {\bibfnamefont {A.}~\bibnamefont {Retzker}},\ }\href {\doibase
		10.1103/PhysRevLett.105.075701} {\bibfield  {journal} {\bibinfo  {journal}
			{Phys. Rev. Lett.}\ }\textbf {\bibinfo {volume} {105}},\ \bibinfo {pages}
		{075701} (\bibinfo {year} {2010})}\BibitemShut {NoStop}%
	\bibitem [{\citenamefont {Chiara}\ \emph {et~al.}(2010)\citenamefont {Chiara},
		\citenamefont {del Campo}, \citenamefont {Morigi}, \citenamefont {Plenio},\
		and\ \citenamefont {Retzker}}]{ions2}%
	\BibitemOpen
	\bibfield  {author} {\bibinfo {author} {\bibfnamefont {G.~D.}\ \bibnamefont
			{Chiara}}, \bibinfo {author} {\bibfnamefont {A.}~\bibnamefont {del Campo}},
		\bibinfo {author} {\bibfnamefont {G.}~\bibnamefont {Morigi}}, \bibinfo
		{author} {\bibfnamefont {M.~B.}\ \bibnamefont {Plenio}}, \ and\ \bibinfo
		{author} {\bibfnamefont {A.}~\bibnamefont {Retzker}},\ }\href
	{http://stacks.iop.org/1367-2630/12/i=11/a=115003} {\bibfield  {journal}
		{\bibinfo  {journal} {New J. Phys.}\ }\textbf {\bibinfo {volume} {12}},\
		\bibinfo {pages} {115003} (\bibinfo {year} {2010})}\BibitemShut {NoStop}%
	\bibitem [{\citenamefont {Witkowska}\ \emph {et~al.}(2011)\citenamefont
		{Witkowska}, \citenamefont {Deuar}, \citenamefont {Gajda},\ and\
		\citenamefont {Rz\k{a}\ifmmode~\dot{z}\else \.{z}\fi{}ewski}}]{WDGR11}%
	\BibitemOpen
	\bibfield  {author} {\bibinfo {author} {\bibfnamefont {E.}~\bibnamefont
			{Witkowska}}, \bibinfo {author} {\bibfnamefont {P.}~\bibnamefont {Deuar}},
		\bibinfo {author} {\bibfnamefont {M.}~\bibnamefont {Gajda}}, \ and\ \bibinfo
		{author} {\bibfnamefont {K.}~\bibnamefont {Rz\k{a}\ifmmode~\dot{z}\else
				\.{z}\fi{}ewski}},\ }\href {\doibase 10.1103/PhysRevLett.106.135301}
	{\bibfield  {journal} {\bibinfo  {journal} {Phys. Rev. Lett.}\ }\textbf
		{\bibinfo {volume} {106}},\ \bibinfo {pages} {135301} (\bibinfo {year}
		{2011})}\BibitemShut {NoStop}%
	\bibitem [{\citenamefont {Saito}\ \emph {et~al.}(2007)\citenamefont {Saito},
		\citenamefont {Kawaguchi},\ and\ \citenamefont {Ueda}}]{dkzm1}%
	\BibitemOpen
	\bibfield  {author} {\bibinfo {author} {\bibfnamefont {H.}~\bibnamefont
			{Saito}}, \bibinfo {author} {\bibfnamefont {Y.}~\bibnamefont {Kawaguchi}}, \
		and\ \bibinfo {author} {\bibfnamefont {M.}~\bibnamefont {Ueda}},\ }\href
	{\doibase 10.1103/PhysRevA.76.043613} {\bibfield  {journal} {\bibinfo
			{journal} {Phys. Rev. A}\ }\textbf {\bibinfo {volume} {76}},\ \bibinfo
		{pages} {043613} (\bibinfo {year} {2007})}\BibitemShut {NoStop}%
	\bibitem [{\citenamefont {Dziarmaga}\ \emph {et~al.}(2008)\citenamefont
		{Dziarmaga}, \citenamefont {Meisner},\ and\ \citenamefont {Zurek}}]{dkzm2}%
	\BibitemOpen
	\bibfield  {author} {\bibinfo {author} {\bibfnamefont {J.}~\bibnamefont
			{Dziarmaga}}, \bibinfo {author} {\bibfnamefont {J.}~\bibnamefont {Meisner}},
		\ and\ \bibinfo {author} {\bibfnamefont {W.~H.}\ \bibnamefont {Zurek}},\
	}\href {\doibase 10.1103/PhysRevLett.101.115701} {\bibfield  {journal}
	{\bibinfo  {journal} {Phys. Rev. Lett.}\ }\textbf {\bibinfo {volume} {101}},\
	\bibinfo {pages} {115701} (\bibinfo {year} {2008})}\BibitemShut {NoStop}%
\bibitem [{\citenamefont {Nigmatullin}\ \emph {et~al.}(2016)\citenamefont
	{Nigmatullin}, \citenamefont {del Campo}, \citenamefont {De~Chiara},
	\citenamefont {Morigi}, \citenamefont {Plenio},\ and\ \citenamefont
	{Retzker}}]{Nigmatullin11}%
\BibitemOpen
\bibfield  {author} {\bibinfo {author} {\bibfnamefont {R.}~\bibnamefont
		{Nigmatullin}}, \bibinfo {author} {\bibfnamefont {A.}~\bibnamefont {del
			Campo}}, \bibinfo {author} {\bibfnamefont {G.}~\bibnamefont {De~Chiara}},
	\bibinfo {author} {\bibfnamefont {G.}~\bibnamefont {Morigi}}, \bibinfo
	{author} {\bibfnamefont {M.~B.}\ \bibnamefont {Plenio}}, \ and\ \bibinfo
	{author} {\bibfnamefont {A.}~\bibnamefont {Retzker}},\ }\href {\doibase
	10.1103/PhysRevB.93.014106} {\bibfield  {journal} {\bibinfo  {journal} {Phys.
			Rev. B}\ }\textbf {\bibinfo {volume} {93}},\ \bibinfo {pages} {014106}
	(\bibinfo {year} {2016})}\BibitemShut {NoStop}%
\bibitem [{\citenamefont {Das}\ \emph {et~al.}(2012)\citenamefont {Das},
	\citenamefont {Sabbatini},\ and\ \citenamefont {Zurek}}]{DSZ12}%
\BibitemOpen
\bibfield  {author} {\bibinfo {author} {\bibfnamefont {A.}~\bibnamefont
		{Das}}, \bibinfo {author} {\bibfnamefont {J.}~\bibnamefont {Sabbatini}}, \
	and\ \bibinfo {author} {\bibfnamefont {W.~H.}\ \bibnamefont {Zurek}},\ }\href
{\doibase doi:10.1038/srep00352} {\bibfield  {journal} {\bibinfo  {journal}
		{Sci. Rep.}\ }\textbf {\bibinfo {volume} {2}},\ \bibinfo {pages} {352}
	(\bibinfo {year} {2012})}\BibitemShut {NoStop}%
\bibitem [{\citenamefont {Chesler}\ \emph {et~al.}(2015)\citenamefont
	{Chesler}, \citenamefont {Garc\'{\i}a-Garc\'{\i}a},\ and\ \citenamefont
	{Liu}}]{holo2}%
\BibitemOpen
\bibfield  {author} {\bibinfo {author} {\bibfnamefont {P.~M.}\ \bibnamefont
		{Chesler}}, \bibinfo {author} {\bibfnamefont {A.~M.}\ \bibnamefont
		{Garc\'{\i}a-Garc\'{\i}a}}, \ and\ \bibinfo {author} {\bibfnamefont
		{H.}~\bibnamefont {Liu}},\ }\href {\doibase 10.1103/PhysRevX.5.021015}
{\bibfield  {journal} {\bibinfo  {journal} {Phys. Rev. X}\ }\textbf {\bibinfo
		{volume} {5}},\ \bibinfo {pages} {021015} (\bibinfo {year}
	{2015})}\BibitemShut {NoStop}%
\bibitem [{\citenamefont {Sonner}\ \emph {et~al.}(2015)\citenamefont {Sonner},
	\citenamefont {del Campo},\ and\ \citenamefont {Zurek}}]{Sonner}%
\BibitemOpen
\bibfield  {author} {\bibinfo {author} {\bibfnamefont {J.}~\bibnamefont
		{Sonner}}, \bibinfo {author} {\bibfnamefont {A.}~\bibnamefont {del Campo}}, \
	and\ \bibinfo {author} {\bibfnamefont {W.~H.}\ \bibnamefont {Zurek}},\ }\href
{http://dx.doi.org/10.1038/ncomms8406} {\bibfield  {journal} {\bibinfo
		{journal} {Nature Communications}\ }\textbf {\bibinfo {volume} {6}},\
	\bibinfo {pages} {7406} (\bibinfo {year} {2015})},\ \bibinfo {note}
{article}\BibitemShut {NoStop}%
\bibitem [{\citenamefont {Francuz}\ \emph {et~al.}(2016)\citenamefont
	{Francuz}, \citenamefont {Dziarmaga}, \citenamefont {Gardas},\ and\
	\citenamefont {Zurek}}]{Francuzetal}%
\BibitemOpen
\bibfield  {author} {\bibinfo {author} {\bibfnamefont {A.}~\bibnamefont
		{Francuz}}, \bibinfo {author} {\bibfnamefont {J.}~\bibnamefont {Dziarmaga}},
	\bibinfo {author} {\bibfnamefont {B.}~\bibnamefont {Gardas}}, \ and\ \bibinfo
	{author} {\bibfnamefont {W.~H.}\ \bibnamefont {Zurek}},\ }\href {\doibase
	10.1103/PhysRevB.93.075134} {\bibfield  {journal} {\bibinfo  {journal} {Phys.
			Rev. B}\ }\textbf {\bibinfo {volume} {93}},\ \bibinfo {pages} {075134}
	(\bibinfo {year} {2016})}\BibitemShut {NoStop}%
\bibitem [{\citenamefont {Chuang}\ \emph {et~al.}(1991)\citenamefont {Chuang},
	\citenamefont {Durrer}, \citenamefont {Turok},\ and\ \citenamefont
	{Yurke}}]{Chuang91}%
\BibitemOpen
\bibfield  {author} {\bibinfo {author} {\bibfnamefont {I.}~\bibnamefont
		{Chuang}}, \bibinfo {author} {\bibfnamefont {R.}~\bibnamefont {Durrer}},
	\bibinfo {author} {\bibfnamefont {N.}~\bibnamefont {Turok}}, \ and\ \bibinfo
	{author} {\bibfnamefont {B.}~\bibnamefont {Yurke}},\ }\href {\doibase
	10.1126/science.251.4999.1336} {\bibfield  {journal} {\bibinfo  {journal}
		{Science}\ }\textbf {\bibinfo {volume} {251}},\ \bibinfo {pages} {1336}
	(\bibinfo {year} {1991})}\BibitemShut {NoStop}%
\bibitem [{\citenamefont {Bowick}\ \emph {et~al.}(1994)\citenamefont {Bowick},
	\citenamefont {Chandar}, \citenamefont {Schiff},\ and\ \citenamefont
	{Srivastava}}]{Bowick94}%
\BibitemOpen
\bibfield  {author} {\bibinfo {author} {\bibfnamefont {M.~J.}\ \bibnamefont
		{Bowick}}, \bibinfo {author} {\bibfnamefont {L.}~\bibnamefont {Chandar}},
	\bibinfo {author} {\bibfnamefont {E.~A.}\ \bibnamefont {Schiff}}, \ and\
	\bibinfo {author} {\bibfnamefont {A.~M.}\ \bibnamefont {Srivastava}},\ }\href
{\doibase 10.1126/science.263.5149.943} {\bibfield  {journal} {\bibinfo
		{journal} {Science}\ }\textbf {\bibinfo {volume} {263}},\ \bibinfo {pages}
	{943} (\bibinfo {year} {1994})}\BibitemShut {NoStop}%
\bibitem [{\citenamefont {Ruutu}\ \emph {et~al.}(1996)\citenamefont {Ruutu},
	\citenamefont {Eltsov}, \citenamefont {Gill}, \citenamefont {Kibble},
	\citenamefont {Krusius}, \citenamefont {Makhlin}, \citenamefont {Placais},
	\citenamefont {Volovik},\ and\ \citenamefont {Xu}}]{Ruutu96}%
\BibitemOpen
\bibfield  {author} {\bibinfo {author} {\bibfnamefont {V.~M.~H.}\
		\bibnamefont {Ruutu}}, \bibinfo {author} {\bibfnamefont {V.~B.}\ \bibnamefont
		{Eltsov}}, \bibinfo {author} {\bibfnamefont {A.~J.}\ \bibnamefont {Gill}},
	\bibinfo {author} {\bibfnamefont {T.~W.~B.}\ \bibnamefont {Kibble}}, \bibinfo
	{author} {\bibfnamefont {M.}~\bibnamefont {Krusius}}, \bibinfo {author}
	{\bibfnamefont {Y.~G.}\ \bibnamefont {Makhlin}}, \bibinfo {author}
	{\bibfnamefont {B.}~\bibnamefont {Placais}}, \bibinfo {author} {\bibfnamefont
		{G.~E.}\ \bibnamefont {Volovik}}, \ and\ \bibinfo {author} {\bibfnamefont
		{W.}~\bibnamefont {Xu}},\ }\href {\doibase 10.1038/382334a0} {\bibfield
	{journal} {\bibinfo  {journal} {Nature}\ }\textbf {\bibinfo {volume} {382}},\
	\bibinfo {pages} {334} (\bibinfo {year} {1996})}\BibitemShut {NoStop}%
\bibitem [{\citenamefont {Bauerle}\ \emph {et~al.}(1996)\citenamefont
	{Bauerle}, \citenamefont {Bunkov}, \citenamefont {Fisher}, \citenamefont
	{Godfrin},\ and\ \citenamefont {Pickett}}]{Bauerle96}%
\BibitemOpen
\bibfield  {author} {\bibinfo {author} {\bibfnamefont {C.}~\bibnamefont
		{Bauerle}}, \bibinfo {author} {\bibfnamefont {Y.~M.}\ \bibnamefont {Bunkov}},
	\bibinfo {author} {\bibfnamefont {S.~N.}\ \bibnamefont {Fisher}}, \bibinfo
	{author} {\bibfnamefont {H.}~\bibnamefont {Godfrin}}, \ and\ \bibinfo
	{author} {\bibfnamefont {G.~R.}\ \bibnamefont {Pickett}},\ }\href {\doibase
	10.1038/382332a0} {\bibfield  {journal} {\bibinfo  {journal} {Nature}\
	}\textbf {\bibinfo {volume} {382}},\ \bibinfo {pages} {332} (\bibinfo {year}
	{1996})}\BibitemShut {NoStop}%
\bibitem [{\citenamefont {Carmi}\ \emph {et~al.}(2000)\citenamefont {Carmi},
	\citenamefont {Polturak},\ and\ \citenamefont {Koren}}]{Carmi00}%
\BibitemOpen
\bibfield  {author} {\bibinfo {author} {\bibfnamefont {R.}~\bibnamefont
		{Carmi}}, \bibinfo {author} {\bibfnamefont {E.}~\bibnamefont {Polturak}}, \
	and\ \bibinfo {author} {\bibfnamefont {G.}~\bibnamefont {Koren}},\ }\href
{\doibase 10.1103/PhysRevLett.84.4966} {\bibfield  {journal} {\bibinfo
		{journal} {Phys. Rev. Lett.}\ }\textbf {\bibinfo {volume} {84}},\ \bibinfo
	{pages} {4966} (\bibinfo {year} {2000})}\BibitemShut {NoStop}%
\bibitem [{\citenamefont {Monaco}\ \emph {et~al.}(2002)\citenamefont {Monaco},
	\citenamefont {Mygind},\ and\ \citenamefont {Rivers}}]{Monaco02}%
\BibitemOpen
\bibfield  {author} {\bibinfo {author} {\bibfnamefont {R.}~\bibnamefont
		{Monaco}}, \bibinfo {author} {\bibfnamefont {J.}~\bibnamefont {Mygind}}, \
	and\ \bibinfo {author} {\bibfnamefont {R.~J.}\ \bibnamefont {Rivers}},\
}\href {\doibase 10.1103/PhysRevLett.89.080603} {\bibfield  {journal}
{\bibinfo  {journal} {Phys. Rev. Lett.}\ }\textbf {\bibinfo {volume} {89}},\
\bibinfo {pages} {080603} (\bibinfo {year} {2002})}\BibitemShut {NoStop}%
\bibitem [{\citenamefont {Monaco}\ \emph {et~al.}(2009)\citenamefont {Monaco},
	\citenamefont {Mygind}, \citenamefont {Rivers},\ and\ \citenamefont
	{Koshelets}}]{Monaco09}%
\BibitemOpen
\bibfield  {author} {\bibinfo {author} {\bibfnamefont {R.}~\bibnamefont
		{Monaco}}, \bibinfo {author} {\bibfnamefont {J.}~\bibnamefont {Mygind}},
	\bibinfo {author} {\bibfnamefont {R.~J.}\ \bibnamefont {Rivers}}, \ and\
	\bibinfo {author} {\bibfnamefont {V.~P.}\ \bibnamefont {Koshelets}},\ }\href
{\doibase 10.1103/PhysRevB.80.180501} {\bibfield  {journal} {\bibinfo
		{journal} {Phys. Rev. B}\ }\textbf {\bibinfo {volume} {80}},\ \bibinfo
	{pages} {180501} (\bibinfo {year} {2009})}\BibitemShut {NoStop}%
\bibitem [{\citenamefont {Maniv}\ \emph {et~al.}(2003)\citenamefont {Maniv},
	\citenamefont {Polturak},\ and\ \citenamefont {Koren}}]{Maniv03}%
\BibitemOpen
\bibfield  {author} {\bibinfo {author} {\bibfnamefont {A.}~\bibnamefont
		{Maniv}}, \bibinfo {author} {\bibfnamefont {E.}~\bibnamefont {Polturak}}, \
	and\ \bibinfo {author} {\bibfnamefont {G.}~\bibnamefont {Koren}},\ }\href
{\doibase 10.1103/PhysRevLett.91.197001} {\bibfield  {journal} {\bibinfo
		{journal} {Phys. Rev. Lett.}\ }\textbf {\bibinfo {volume} {91}},\ \bibinfo
	{pages} {197001} (\bibinfo {year} {2003})}\BibitemShut {NoStop}%
\bibitem [{\citenamefont {Sadler}\ \emph {et~al.}(2006)\citenamefont {Sadler},
	\citenamefont {Higbie}, \citenamefont {Leslie}, \citenamefont
	{Vengalattore},\ and\ \citenamefont {Stamper-Kurn}}]{Sadler06}%
\BibitemOpen
\bibfield  {author} {\bibinfo {author} {\bibfnamefont {L.~E.}\ \bibnamefont
		{Sadler}}, \bibinfo {author} {\bibfnamefont {J.~M.}\ \bibnamefont {Higbie}},
	\bibinfo {author} {\bibfnamefont {S.~R.}\ \bibnamefont {Leslie}}, \bibinfo
	{author} {\bibfnamefont {M.}~\bibnamefont {Vengalattore}}, \ and\ \bibinfo
	{author} {\bibfnamefont {D.~M.}\ \bibnamefont {Stamper-Kurn}},\ }\href
{\doibase 10.1038/nature05094} {\bibfield  {journal} {\bibinfo  {journal}
		{Nature}\ }\textbf {\bibinfo {volume} {443}},\ \bibinfo {pages} {312}
	(\bibinfo {year} {2006})}\BibitemShut {NoStop}%
\bibitem [{\citenamefont {Golubchik}\ \emph {et~al.}(2010)\citenamefont
	{Golubchik}, \citenamefont {Polturak},\ and\ \citenamefont
	{Koren}}]{Golubchik10}%
\BibitemOpen
\bibfield  {author} {\bibinfo {author} {\bibfnamefont {D.}~\bibnamefont
		{Golubchik}}, \bibinfo {author} {\bibfnamefont {E.}~\bibnamefont {Polturak}},
	\ and\ \bibinfo {author} {\bibfnamefont {G.}~\bibnamefont {Koren}},\ }\href
{\doibase 10.1103/PhysRevLett.104.247002} {\bibfield  {journal} {\bibinfo
		{journal} {Phys. Rev. Lett.}\ }\textbf {\bibinfo {volume} {104}},\ \bibinfo
	{pages} {247002} (\bibinfo {year} {2010})}\BibitemShut {NoStop}%
\bibitem [{\citenamefont {Chae}\ \emph {et~al.}(2012)\citenamefont {Chae},
	\citenamefont {Lee}, \citenamefont {Horibe}, \citenamefont {Tanimura},
	\citenamefont {Mori}, \citenamefont {Gao}, \citenamefont {Carr},\ and\
	\citenamefont {Cheong}}]{Chae12}%
\BibitemOpen
\bibfield  {author} {\bibinfo {author} {\bibfnamefont {S.~C.}\ \bibnamefont
		{Chae}}, \bibinfo {author} {\bibfnamefont {N.}~\bibnamefont {Lee}}, \bibinfo
	{author} {\bibfnamefont {Y.}~\bibnamefont {Horibe}}, \bibinfo {author}
	{\bibfnamefont {M.}~\bibnamefont {Tanimura}}, \bibinfo {author}
	{\bibfnamefont {S.}~\bibnamefont {Mori}}, \bibinfo {author} {\bibfnamefont
		{B.}~\bibnamefont {Gao}}, \bibinfo {author} {\bibfnamefont {S.}~\bibnamefont
		{Carr}}, \ and\ \bibinfo {author} {\bibfnamefont {S.-W.}\ \bibnamefont
		{Cheong}},\ }\href {\doibase 10.1103/PhysRevLett.108.167603} {\bibfield
	{journal} {\bibinfo  {journal} {Phys. Rev. Lett.}\ }\textbf {\bibinfo
		{volume} {108}},\ \bibinfo {pages} {167603} (\bibinfo {year}
	{2012})}\BibitemShut {NoStop}%
\bibitem [{\citenamefont {Griffin}\ \emph {et~al.}(2012)\citenamefont
	{Griffin}, \citenamefont {Lilienblum}, \citenamefont {Delaney}, \citenamefont
	{Kumagai}, \citenamefont {Fiebig},\ and\ \citenamefont
	{Spaldin}}]{Griffin12}%
\BibitemOpen
\bibfield  {author} {\bibinfo {author} {\bibfnamefont {S.~M.}\ \bibnamefont
		{Griffin}}, \bibinfo {author} {\bibfnamefont {M.}~\bibnamefont {Lilienblum}},
	\bibinfo {author} {\bibfnamefont {K.~T.}\ \bibnamefont {Delaney}}, \bibinfo
	{author} {\bibfnamefont {Y.}~\bibnamefont {Kumagai}}, \bibinfo {author}
	{\bibfnamefont {M.}~\bibnamefont {Fiebig}}, \ and\ \bibinfo {author}
	{\bibfnamefont {N.~A.}\ \bibnamefont {Spaldin}},\ }\href {\doibase
	10.1103/PhysRevX.2.041022} {\bibfield  {journal} {\bibinfo  {journal} {Phys.
			Rev. X}\ }\textbf {\bibinfo {volume} {2}},\ \bibinfo {pages} {041022}
	(\bibinfo {year} {2012})}\BibitemShut {NoStop}%
\bibitem [{\citenamefont {Mielenz}\ \emph {et~al.}(2013)\citenamefont
	{Mielenz}, \citenamefont {Brox}, \citenamefont {Kahra}, \citenamefont
	{Leschhorn}, \citenamefont {Albert}, \citenamefont {Schaetz}, \citenamefont
	{Landa},\ and\ \citenamefont {Reznik}}]{Schaetz13}%
\BibitemOpen
\bibfield  {author} {\bibinfo {author} {\bibfnamefont {M.}~\bibnamefont
		{Mielenz}}, \bibinfo {author} {\bibfnamefont {J.}~\bibnamefont {Brox}},
	\bibinfo {author} {\bibfnamefont {S.}~\bibnamefont {Kahra}}, \bibinfo
	{author} {\bibfnamefont {G.}~\bibnamefont {Leschhorn}}, \bibinfo {author}
	{\bibfnamefont {M.}~\bibnamefont {Albert}}, \bibinfo {author} {\bibfnamefont
		{T.}~\bibnamefont {Schaetz}}, \bibinfo {author} {\bibfnamefont
		{H.}~\bibnamefont {Landa}}, \ and\ \bibinfo {author} {\bibfnamefont
		{B.}~\bibnamefont {Reznik}},\ }\href {\doibase
	10.1103/PhysRevLett.110.133004} {\bibfield  {journal} {\bibinfo  {journal}
		{Phys. Rev. Lett.}\ }\textbf {\bibinfo {volume} {110}},\ \bibinfo {pages}
	{133004} (\bibinfo {year} {2013})}\BibitemShut {NoStop}%
\bibitem [{\citenamefont {Ejtemaee}\ and\ \citenamefont {Haljan}(2013)}]{EH13}%
\BibitemOpen
\bibfield  {author} {\bibinfo {author} {\bibfnamefont {S.}~\bibnamefont
		{Ejtemaee}}\ and\ \bibinfo {author} {\bibfnamefont {P.~C.}\ \bibnamefont
		{Haljan}},\ }\href {\doibase 10.1103/PhysRevA.87.051401} {\bibfield
	{journal} {\bibinfo  {journal} {Phys. Rev. A}\ }\textbf {\bibinfo {volume}
		{87}},\ \bibinfo {pages} {051401} (\bibinfo {year} {2013})}\BibitemShut
{NoStop}%
\bibitem [{\citenamefont {S.~Ulm}\ \emph {et~al.}(2013)\citenamefont {S.~Ulm},
	\citenamefont {Ro{\ss}nagel}, \citenamefont {Jacob}, \citenamefont
	{Deg{\"u}nther}, \citenamefont {Dawkins}, \citenamefont {Poschinger},
	\citenamefont {Nigmatullin}, \citenamefont {Retzker}, \citenamefont {Plenio},
	\citenamefont {Schmidt-Kaler},\ and\ \citenamefont {Singer}}]{Ulm13}%
\BibitemOpen
\bibfield  {author} {\bibinfo {author} {\bibfnamefont {S.}~\bibnamefont
		{S.~Ulm}}, \bibinfo {author} {\bibfnamefont {J.}~\bibnamefont
		{Ro{\ss}nagel}}, \bibinfo {author} {\bibfnamefont {G.}~\bibnamefont {Jacob}},
	\bibinfo {author} {\bibfnamefont {C.}~\bibnamefont {Deg{\"u}nther}}, \bibinfo
	{author} {\bibfnamefont {S.~T.}\ \bibnamefont {Dawkins}}, \bibinfo {author}
	{\bibfnamefont {U.~G.}\ \bibnamefont {Poschinger}}, \bibinfo {author}
	{\bibfnamefont {R.}~\bibnamefont {Nigmatullin}}, \bibinfo {author}
	{\bibfnamefont {A.}~\bibnamefont {Retzker}}, \bibinfo {author} {\bibfnamefont
		{M.~B.}\ \bibnamefont {Plenio}}, \bibinfo {author} {\bibfnamefont
		{F.}~\bibnamefont {Schmidt-Kaler}}, \ and\ \bibinfo {author} {\bibfnamefont
		{K.}~\bibnamefont {Singer}},\ }\href {\doibase doi:10.1038/ncomms3290}
{\bibfield  {journal} {\bibinfo  {journal} {Nat. Commun.}\ }\textbf {\bibinfo
		{volume} {4}},\ \bibinfo {pages} {2290} (\bibinfo {year} {2013})}\BibitemShut
{NoStop}%
\bibitem [{\citenamefont {Pyka}\ \emph {et~al.}(2013)\citenamefont {Pyka},
	\citenamefont {Keller}, \citenamefont {Partner}, \citenamefont {Nigmatullin},
	\citenamefont {Burgermeister}, \citenamefont {Meier}, \citenamefont
	{Kuhlmann}, \citenamefont {Retzker}, \citenamefont {Plenio}, \citenamefont
	{Zurek}, \citenamefont {del Campo},\ and\ \citenamefont
	{Mehlst{\"a}ubler}}]{Tanja13}%
\BibitemOpen
\bibfield  {author} {\bibinfo {author} {\bibfnamefont {K.}~\bibnamefont
		{Pyka}}, \bibinfo {author} {\bibfnamefont {J.}~\bibnamefont {Keller}},
	\bibinfo {author} {\bibfnamefont {H.~L.}\ \bibnamefont {Partner}}, \bibinfo
	{author} {\bibfnamefont {R.}~\bibnamefont {Nigmatullin}}, \bibinfo {author}
	{\bibfnamefont {T.}~\bibnamefont {Burgermeister}}, \bibinfo {author}
	{\bibfnamefont {D.~M.}\ \bibnamefont {Meier}}, \bibinfo {author}
	{\bibfnamefont {K.}~\bibnamefont {Kuhlmann}}, \bibinfo {author}
	{\bibfnamefont {A.}~\bibnamefont {Retzker}}, \bibinfo {author} {\bibfnamefont
		{M.~B.}\ \bibnamefont {Plenio}}, \bibinfo {author} {\bibfnamefont {W.~H.}\
		\bibnamefont {Zurek}}, \bibinfo {author} {\bibfnamefont {A.}~\bibnamefont
		{del Campo}}, \ and\ \bibinfo {author} {\bibfnamefont {T.~E.}\ \bibnamefont
		{Mehlst{\"a}ubler}},\ }\href {http://dx.doi.org/10.1038/ncomms3291}
{\bibfield  {journal} {\bibinfo  {journal} {Nat. Commun.}\ }\textbf {\bibinfo
		{volume} {4}},\ \bibinfo {pages} {2291} (\bibinfo {year} {2013})}\BibitemShut
{NoStop}%
\bibitem [{\citenamefont {Weiler}\ \emph {et~al.}(2008)\citenamefont {Weiler},
	\citenamefont {Neely}, \citenamefont {Scherer}, \citenamefont {Bradley},
	\citenamefont {Davis},\ and\ \citenamefont {Anderson}}]{Anderson08}%
\BibitemOpen
\bibfield  {author} {\bibinfo {author} {\bibfnamefont {C.~N.}\ \bibnamefont
		{Weiler}}, \bibinfo {author} {\bibfnamefont {T.~W.}\ \bibnamefont {Neely}},
	\bibinfo {author} {\bibfnamefont {D.~R.}\ \bibnamefont {Scherer}}, \bibinfo
	{author} {\bibfnamefont {A.~S.}\ \bibnamefont {Bradley}}, \bibinfo {author}
	{\bibfnamefont {M.~J.}\ \bibnamefont {Davis}}, \ and\ \bibinfo {author}
	{\bibfnamefont {B.~P.}\ \bibnamefont {Anderson}},\ }\href {\doibase
	10.1038/nature07334} {\bibfield  {journal} {\bibinfo  {journal} {Nature}\
	}\textbf {\bibinfo {volume} {455}},\ \bibinfo {pages} {948} (\bibinfo {year}
	{2008})}\BibitemShut {NoStop}%
\bibitem [{\citenamefont {Lamporesi}\ \emph {et~al.}(2013)\citenamefont
	{Lamporesi}, \citenamefont {Donadello}, \citenamefont {Serafini},
	\citenamefont {Dalfovo},\ and\ \citenamefont {Ferrari}}]{Lamporesi13}%
\BibitemOpen
\bibfield  {author} {\bibinfo {author} {\bibfnamefont {G.}~\bibnamefont
		{Lamporesi}}, \bibinfo {author} {\bibfnamefont {S.}~\bibnamefont
		{Donadello}}, \bibinfo {author} {\bibfnamefont {S.}~\bibnamefont {Serafini}},
	\bibinfo {author} {\bibfnamefont {F.}~\bibnamefont {Dalfovo}}, \ and\
	\bibinfo {author} {\bibfnamefont {G.}~\bibnamefont {Ferrari}},\ }\href
{http://dx.doi.org/10.1038/nphys2734} {\bibfield  {journal} {\bibinfo
		{journal} {Nat. Phys.}\ }\textbf {\bibinfo {volume} {9}},\ \bibinfo {pages}
	{656} (\bibinfo {year} {2013})}\BibitemShut {NoStop}%
\bibitem [{\citenamefont {Corman}\ \emph {et~al.}(2014)\citenamefont {Corman},
	\citenamefont {Chomaz}, \citenamefont {Bienaim\'e}, \citenamefont
	{Desbuquois}, \citenamefont {Weitenberg}, \citenamefont {Nascimb\`ene},
	\citenamefont {Dalibard},\ and\ \citenamefont
	{Beugnon}}]{DalibardSupercurrents}%
\BibitemOpen
\bibfield  {author} {\bibinfo {author} {\bibfnamefont {L.}~\bibnamefont
		{Corman}}, \bibinfo {author} {\bibfnamefont {L.}~\bibnamefont {Chomaz}},
	\bibinfo {author} {\bibfnamefont {T.}~\bibnamefont {Bienaim\'e}}, \bibinfo
	{author} {\bibfnamefont {R.}~\bibnamefont {Desbuquois}}, \bibinfo {author}
	{\bibfnamefont {C.}~\bibnamefont {Weitenberg}}, \bibinfo {author}
	{\bibfnamefont {S.}~\bibnamefont {Nascimb\`ene}}, \bibinfo {author}
	{\bibfnamefont {J.}~\bibnamefont {Dalibard}}, \ and\ \bibinfo {author}
	{\bibfnamefont {J.}~\bibnamefont {Beugnon}},\ }\href {\doibase
	10.1103/PhysRevLett.113.135302} {\bibfield  {journal} {\bibinfo  {journal}
		{Phys. Rev. Lett.}\ }\textbf {\bibinfo {volume} {113}},\ \bibinfo {pages}
	{135302} (\bibinfo {year} {2014})}\BibitemShut {NoStop}%
\bibitem [{\citenamefont {Chomaz}\ \emph {et~al.}(2015)\citenamefont {Chomaz},
	\citenamefont {Corman}, \citenamefont {Bienaim{\'e}}, \citenamefont
	{Desbuquois}, \citenamefont {Weitenberg}, \citenamefont {Nascimb{\`e}ne},
	\citenamefont {Beugnon},\ and\ \citenamefont {Dalibard}}]{DalibardCoherence}%
\BibitemOpen
\bibfield  {author} {\bibinfo {author} {\bibfnamefont {L.}~\bibnamefont
		{Chomaz}}, \bibinfo {author} {\bibfnamefont {L.}~\bibnamefont {Corman}},
	\bibinfo {author} {\bibfnamefont {T.}~\bibnamefont {Bienaim{\'e}}}, \bibinfo
	{author} {\bibfnamefont {R.}~\bibnamefont {Desbuquois}}, \bibinfo {author}
	{\bibfnamefont {C.}~\bibnamefont {Weitenberg}}, \bibinfo {author}
	{\bibfnamefont {S.}~\bibnamefont {Nascimb{\`e}ne}}, \bibinfo {author}
	{\bibfnamefont {J.}~\bibnamefont {Beugnon}}, \ and\ \bibinfo {author}
	{\bibfnamefont {J.}~\bibnamefont {Dalibard}},\ }\href
{http://dx.doi.org/10.1038/ncomms7162} {\bibfield  {journal} {\bibinfo
		{journal} {Nat. Commun.}\ }\textbf {\bibinfo {volume} {6}},\ \bibinfo {pages}
	{6162} (\bibinfo {year} {2015})}\BibitemShut {NoStop}%
\bibitem [{\citenamefont {Lin}\ \emph {et~al.}(2014)\citenamefont {Lin},
	\citenamefont {Wang}, \citenamefont {Kamiya}, \citenamefont {Chern},
	\citenamefont {Fan}, \citenamefont {Fan}, \citenamefont {Casas},
	\citenamefont {Liu}, \citenamefont {Kiryukhin}, \citenamefont {Zurek},
	\citenamefont {Batista},\ and\ \citenamefont {Cheong}}]{ferroelectrics}%
\BibitemOpen
\bibfield  {author} {\bibinfo {author} {\bibfnamefont {S.-Z.}\ \bibnamefont
		{Lin}}, \bibinfo {author} {\bibfnamefont {X.}~\bibnamefont {Wang}}, \bibinfo
	{author} {\bibfnamefont {Y.}~\bibnamefont {Kamiya}}, \bibinfo {author}
	{\bibfnamefont {G.-W.}\ \bibnamefont {Chern}}, \bibinfo {author}
	{\bibfnamefont {F.}~\bibnamefont {Fan}}, \bibinfo {author} {\bibfnamefont
		{D.}~\bibnamefont {Fan}}, \bibinfo {author} {\bibfnamefont {B.}~\bibnamefont
		{Casas}}, \bibinfo {author} {\bibfnamefont {Y.}~\bibnamefont {Liu}}, \bibinfo
	{author} {\bibfnamefont {V.}~\bibnamefont {Kiryukhin}}, \bibinfo {author}
	{\bibfnamefont {W.~H.}\ \bibnamefont {Zurek}}, \bibinfo {author}
	{\bibfnamefont {C.~D.}\ \bibnamefont {Batista}}, \ and\ \bibinfo {author}
	{\bibfnamefont {S.-W.}\ \bibnamefont {Cheong}},\ }\href
{http://dx.doi.org/10.1038/nphys3142} {\bibfield  {journal} {\bibinfo
		{journal} {Nat. Phys.}\ }\textbf {\bibinfo {volume} {10}},\ \bibinfo {pages}
	{970} (\bibinfo {year} {2014})}\BibitemShut {NoStop}%
\bibitem [{\citenamefont {Navon}\ \emph {et~al.}(2015)\citenamefont {Navon},
	\citenamefont {Gaunt}, \citenamefont {Smith},\ and\ \citenamefont
	{Hadzibabic}}]{Hadzibabic}%
\BibitemOpen
\bibfield  {author} {\bibinfo {author} {\bibfnamefont {N.}~\bibnamefont
		{Navon}}, \bibinfo {author} {\bibfnamefont {A.~L.}\ \bibnamefont {Gaunt}},
	\bibinfo {author} {\bibfnamefont {R.~P.}\ \bibnamefont {Smith}}, \ and\
	\bibinfo {author} {\bibfnamefont {Z.}~\bibnamefont {Hadzibabic}},\ }\href
{\doibase 10.1126/science.1258676} {\bibfield  {journal} {\bibinfo  {journal}
		{Science}\ }\textbf {\bibinfo {volume} {347}},\ \bibinfo {pages} {167}
	(\bibinfo {year} {2015})}\BibitemShut {NoStop}%
\bibitem [{\citenamefont {Beugnon}\ and\ \citenamefont {Navon}(2017)}]{Navon}%
\BibitemOpen
\bibfield  {author} {\bibinfo {author} {\bibfnamefont {J.}~\bibnamefont
		{Beugnon}}\ and\ \bibinfo {author} {\bibfnamefont {N.}~\bibnamefont
		{Navon}},\ }\href {http://stacks.iop.org/0953-4075/50/i=2/a=022002}
{\bibfield  {journal} {\bibinfo  {journal} {J. Phys. B: At. Mol. Opt. Phys.}\
	}\textbf {\bibinfo {volume} {50}},\ \bibinfo {pages} {022002} (\bibinfo
	{year} {2017})}\BibitemShut {NoStop}%
\bibitem [{\citenamefont {Anquez}\ \emph {et~al.}(2016)\citenamefont {Anquez},
	\citenamefont {Robbins}, \citenamefont {Bharath}, \citenamefont
	{Boguslawski}, \citenamefont {Hoang},\ and\ \citenamefont
	{Chapman}}]{FerroKZscaling}%
\BibitemOpen
\bibfield  {author} {\bibinfo {author} {\bibfnamefont {M.}~\bibnamefont
		{Anquez}}, \bibinfo {author} {\bibfnamefont {B.~A.}\ \bibnamefont {Robbins}},
	\bibinfo {author} {\bibfnamefont {H.~M.}\ \bibnamefont {Bharath}}, \bibinfo
	{author} {\bibfnamefont {M.}~\bibnamefont {Boguslawski}}, \bibinfo {author}
	{\bibfnamefont {T.~M.}\ \bibnamefont {Hoang}}, \ and\ \bibinfo {author}
	{\bibfnamefont {M.~S.}\ \bibnamefont {Chapman}},\ }\href {\doibase
	10.1103/PhysRevLett.116.155301} {\bibfield  {journal} {\bibinfo  {journal}
		{Phys. Rev. Lett.}\ }\textbf {\bibinfo {volume} {116}},\ \bibinfo {pages}
	{155301} (\bibinfo {year} {2016})}\BibitemShut {NoStop}%
\bibitem [{\citenamefont {Clark}\ \emph {et~al.}(2016)\citenamefont {Clark},
	\citenamefont {Feng},\ and\ \citenamefont {Chin}}]{Chicago}%
\BibitemOpen
\bibfield  {author} {\bibinfo {author} {\bibfnamefont {L.~W.}\ \bibnamefont
		{Clark}}, \bibinfo {author} {\bibfnamefont {L.}~\bibnamefont {Feng}}, \ and\
	\bibinfo {author} {\bibfnamefont {C.}~\bibnamefont {Chin}},\ }\href {\doibase
	10.1126/science.aaf9657} {\bibfield  {journal} {\bibinfo  {journal}
		{Science}\ }\textbf {\bibinfo {volume} {354}},\ \bibinfo {pages} {606}
	(\bibinfo {year} {2016})}\BibitemShut {NoStop}%
\bibitem [{\citenamefont {del Campo}\ \emph {et~al.}(2013)\citenamefont {del
		Campo}, \citenamefont {Kibble},\ and\ \citenamefont {Zurek}}]{DKZ13}%
\BibitemOpen
\bibfield  {author} {\bibinfo {author} {\bibfnamefont {A.}~\bibnamefont {del
			Campo}}, \bibinfo {author} {\bibfnamefont {T.~W.~B.}\ \bibnamefont {Kibble}},
	\ and\ \bibinfo {author} {\bibfnamefont {W.~H.}\ \bibnamefont {Zurek}},\
}\href {http://stacks.iop.org/0953-8984/25/i=40/a=404210} {\bibfield
{journal} {\bibinfo  {journal} {J. Phys. Condens. Matter}\ }\textbf {\bibinfo
	{volume} {25}},\ \bibinfo {pages} {404210} (\bibinfo {year}
{2013})}\BibitemShut {NoStop}%
\bibitem [{\citenamefont {Zurek}(2013)}]{Zurek13}%
\BibitemOpen
\bibfield  {author} {\bibinfo {author} {\bibfnamefont {W.~H.}\ \bibnamefont
		{Zurek}},\ }\href {http://stacks.iop.org/0953-8984/25/i=40/a=404209}
{\bibfield  {journal} {\bibinfo  {journal} {J. Phys. Condens. Matter}\
	}\textbf {\bibinfo {volume} {25}},\ \bibinfo {pages} {404209} (\bibinfo
	{year} {2013})}\BibitemShut {NoStop}%
\bibitem [{\citenamefont {Damski}\ \emph {et~al.}(2011)\citenamefont {Damski},
	\citenamefont {Quan},\ and\ \citenamefont {Zurek}}]{DQZ11}%
\BibitemOpen
\bibfield  {author} {\bibinfo {author} {\bibfnamefont {B.}~\bibnamefont
		{Damski}}, \bibinfo {author} {\bibfnamefont {H.~T.}\ \bibnamefont {Quan}}, \
	and\ \bibinfo {author} {\bibfnamefont {W.~H.}\ \bibnamefont {Zurek}},\ }\href
{\doibase 10.1103/PhysRevA.83.062104} {\bibfield  {journal} {\bibinfo
		{journal} {Phys. Rev. A}\ }\textbf {\bibinfo {volume} {83}},\ \bibinfo
	{pages} {062104} (\bibinfo {year} {2011})}\BibitemShut {NoStop}%
\bibitem [{\citenamefont {Zurek}(2009)}]{Zurek09}%
\BibitemOpen
\bibfield  {author} {\bibinfo {author} {\bibfnamefont {W.~H.}\ \bibnamefont
		{Zurek}},\ }\href {\doibase 10.1103/PhysRevLett.102.105702} {\bibfield
	{journal} {\bibinfo  {journal} {Phys. Rev. Lett.}\ }\textbf {\bibinfo
		{volume} {102}},\ \bibinfo {pages} {105702} (\bibinfo {year}
	{2009})}\BibitemShut {NoStop}%
\bibitem [{\citenamefont {Damski}\ and\ \citenamefont {Zurek}(2010)}]{DZ10}%
\BibitemOpen
\bibfield  {author} {\bibinfo {author} {\bibfnamefont {B.}~\bibnamefont
		{Damski}}\ and\ \bibinfo {author} {\bibfnamefont {W.~H.}\ \bibnamefont
		{Zurek}},\ }\href {\doibase 10.1103/PhysRevLett.104.160404} {\bibfield
	{journal} {\bibinfo  {journal} {Phys. Rev. Lett.}\ }\textbf {\bibinfo
		{volume} {104}},\ \bibinfo {pages} {160404} (\bibinfo {year}
	{2010})}\BibitemShut {NoStop}%
\bibitem [{\citenamefont {Cincio}\ \emph {et~al.}(2007)\citenamefont {Cincio},
	\citenamefont {Dziarmaga}, \citenamefont {Rams},\ and\ \citenamefont
	{Zurek}}]{Cincio}%
\BibitemOpen
\bibfield  {author} {\bibinfo {author} {\bibfnamefont {L.}~\bibnamefont
		{Cincio}}, \bibinfo {author} {\bibfnamefont {J.}~\bibnamefont {Dziarmaga}},
	\bibinfo {author} {\bibfnamefont {M.~M.}\ \bibnamefont {Rams}}, \ and\
	\bibinfo {author} {\bibfnamefont {W.~H.}\ \bibnamefont {Zurek}},\ }\href
{\doibase 10.1103/PhysRevA.75.052321} {\bibfield  {journal} {\bibinfo
		{journal} {Phys. Rev. A}\ }\textbf {\bibinfo {volume} {75}},\ \bibinfo
	{pages} {052321} (\bibinfo {year} {2007})}\BibitemShut {NoStop}%
\bibitem [{\citenamefont {{Editors H. Arod\'z and J. Dziarmaga and W. H.
			Zurek}}(2003)}]{Kibble03}%
\BibitemOpen
\bibfield  {author} {\bibinfo {author} {\bibnamefont {{Editors H. Arod\'z and
				J. Dziarmaga and W. H. Zurek}}},\ }\href {\doibase 10.1007/978-94-007-1029-0}
{\emph {\bibinfo {title} {T. W. B. Kibble in Patterns of Symmetry
			breaking}}}\ (\bibinfo  {publisher} {Kluwer Academic Publishers},\ \bibinfo
{address} {London},\ \bibinfo {year} {2003})\BibitemShut {NoStop}%
\bibitem [{\citenamefont {Kibble}(2007)}]{Kibble07}%
\BibitemOpen
\bibfield  {author} {\bibinfo {author} {\bibfnamefont {T.~W.~B.}\
		\bibnamefont {Kibble}},\ }\href {\doibase
	http://dx.doi.org/10.1063/1.2784684} {\bibfield  {journal} {\bibinfo
		{journal} {Physics Today}\ }\textbf {\bibinfo {volume} {60}},\ \bibinfo
	{pages} {47} (\bibinfo {year} {2007})}\BibitemShut {NoStop}%
\bibitem [{\citenamefont {Dziarmaga}(2010)}]{Dziarmaga10}%
\BibitemOpen
\bibfield  {author} {\bibinfo {author} {\bibfnamefont {J.}~\bibnamefont
		{Dziarmaga}},\ }\href {\doibase 10.1080/00018732.2010.514702} {\bibfield
	{journal} {\bibinfo  {journal} {Adv. Phys.}\ }\textbf {\bibinfo {volume}
		{59}},\ \bibinfo {pages} {1063} (\bibinfo {year} {2010})}\BibitemShut
{NoStop}%
\bibitem [{\citenamefont {Polkovnikov}\ \emph {et~al.}(2011)\citenamefont
	{Polkovnikov}, \citenamefont {Sengupta}, \citenamefont {Silva},\ and\
	\citenamefont {Vengalattore}}]{Polkovnikov11}%
\BibitemOpen
\bibfield  {author} {\bibinfo {author} {\bibfnamefont {A.}~\bibnamefont
		{Polkovnikov}}, \bibinfo {author} {\bibfnamefont {K.}~\bibnamefont
		{Sengupta}}, \bibinfo {author} {\bibfnamefont {A.}~\bibnamefont {Silva}}, \
	and\ \bibinfo {author} {\bibfnamefont {M.}~\bibnamefont {Vengalattore}},\
}\href {\doibase 10.1103/RevModPhys.83.863} {\bibfield  {journal} {\bibinfo
	{journal} {Rev. Mod. Phys.}\ }\textbf {\bibinfo {volume} {83}},\ \bibinfo
{pages} {863} (\bibinfo {year} {2011})}\BibitemShut {NoStop}%
\bibitem [{\citenamefont {del Campo}\ and\ \citenamefont {Zurek}(2014)}]{DZ13}%
\BibitemOpen
\bibfield  {author} {\bibinfo {author} {\bibfnamefont {A.}~\bibnamefont {del
			Campo}}\ and\ \bibinfo {author} {\bibfnamefont {W.~H.}\ \bibnamefont
		{Zurek}},\ }\href {\doibase 10.1142/S0217751X1430018X} {\bibfield  {journal}
	{\bibinfo  {journal} {Int. J. Mod. Phys. A}\ }\textbf {\bibinfo {volume}
		{29}},\ \bibinfo {pages} {1430018} (\bibinfo {year} {2014})}\BibitemShut
{NoStop}%
\bibitem [{\citenamefont {Dziarmaga}\ \emph {et~al.}(2002)\citenamefont
	{Dziarmaga}, \citenamefont {Smerzi}, \citenamefont {Zurek},\ and\
	\citenamefont {Bishop}}]{Bishop}%
\BibitemOpen
\bibfield  {author} {\bibinfo {author} {\bibfnamefont {J.}~\bibnamefont
		{Dziarmaga}}, \bibinfo {author} {\bibfnamefont {A.}~\bibnamefont {Smerzi}},
	\bibinfo {author} {\bibfnamefont {W.~H.}\ \bibnamefont {Zurek}}, \ and\
	\bibinfo {author} {\bibfnamefont {A.~R.}\ \bibnamefont {Bishop}},\ }\href
{\doibase 10.1103/PhysRevLett.88.167001} {\bibfield  {journal} {\bibinfo
		{journal} {Phys. Rev. Lett.}\ }\textbf {\bibinfo {volume} {88}},\ \bibinfo
	{pages} {167001} (\bibinfo {year} {2002})}\BibitemShut {NoStop}%
\bibitem [{\citenamefont {Damski}(2005)}]{Damski2005}%
\BibitemOpen
\bibfield  {author} {\bibinfo {author} {\bibfnamefont {B.}~\bibnamefont
		{Damski}},\ }\href {\doibase 10.1103/PhysRevLett.95.035701} {\bibfield
	{journal} {\bibinfo  {journal} {Phys. Rev. Lett.}\ }\textbf {\bibinfo
		{volume} {95}},\ \bibinfo {pages} {035701} (\bibinfo {year}
	{2005})}\BibitemShut {NoStop}%
\bibitem [{\citenamefont {Zurek}\ \emph {et~al.}(2005)\citenamefont {Zurek},
	\citenamefont {Dorner},\ and\ \citenamefont {Zoller}}]{Dorner2005}%
\BibitemOpen
\bibfield  {author} {\bibinfo {author} {\bibfnamefont {W.~H.}\ \bibnamefont
		{Zurek}}, \bibinfo {author} {\bibfnamefont {U.}~\bibnamefont {Dorner}}, \
	and\ \bibinfo {author} {\bibfnamefont {P.}~\bibnamefont {Zoller}},\ }\href
{\doibase 10.1103/PhysRevLett.95.105701} {\bibfield  {journal} {\bibinfo
		{journal} {Phys. Rev. Lett.}\ }\textbf {\bibinfo {volume} {95}},\ \bibinfo
	{pages} {105701} (\bibinfo {year} {2005})}\BibitemShut {NoStop}%
\bibitem [{\citenamefont {Dziarmaga}(2005)}]{Dziarmaga2005}%
\BibitemOpen
\bibfield  {author} {\bibinfo {author} {\bibfnamefont {J.}~\bibnamefont
		{Dziarmaga}},\ }\href {\doibase 10.1103/PhysRevLett.95.245701} {\bibfield
	{journal} {\bibinfo  {journal} {Phys. Rev. Lett.}\ }\textbf {\bibinfo
		{volume} {95}},\ \bibinfo {pages} {245701} (\bibinfo {year}
	{2005})}\BibitemShut {NoStop}%
\bibitem [{\citenamefont {Polkovnikov}(2005)}]{Polkovnikov2005}%
\BibitemOpen
\bibfield  {author} {\bibinfo {author} {\bibfnamefont {A.}~\bibnamefont
		{Polkovnikov}},\ }\href {\doibase 10.1103/PhysRevB.72.161201} {\bibfield
	{journal} {\bibinfo  {journal} {Phys. Rev. B}\ }\textbf {\bibinfo {volume}
		{72}},\ \bibinfo {pages} {161201} (\bibinfo {year} {2005})}\BibitemShut
{NoStop}%
\bibitem [{\citenamefont {Mukherjee}\ \emph {et~al.}(2007)\citenamefont
	{Mukherjee}, \citenamefont {Divakaran}, \citenamefont {Dutta},\ and\
	\citenamefont {Sen}}]{ind1}%
\BibitemOpen
\bibfield  {author} {\bibinfo {author} {\bibfnamefont {V.}~\bibnamefont
		{Mukherjee}}, \bibinfo {author} {\bibfnamefont {U.}~\bibnamefont
		{Divakaran}}, \bibinfo {author} {\bibfnamefont {A.}~\bibnamefont {Dutta}}, \
	and\ \bibinfo {author} {\bibfnamefont {D.}~\bibnamefont {Sen}},\ }\href
{\doibase 10.1103/PhysRevB.76.174303} {\bibfield  {journal} {\bibinfo
		{journal} {Phys. Rev. B}\ }\textbf {\bibinfo {volume} {76}},\ \bibinfo
	{pages} {174303} (\bibinfo {year} {2007})}\BibitemShut {NoStop}%
\bibitem [{\citenamefont {Divakaran}\ \emph {et~al.}(2009)\citenamefont
	{Divakaran}, \citenamefont {Mukherjee}, \citenamefont {Dutta},\ and\
	\citenamefont {Sen}}]{ind2}%
\BibitemOpen
\bibfield  {author} {\bibinfo {author} {\bibfnamefont {U.}~\bibnamefont
		{Divakaran}}, \bibinfo {author} {\bibfnamefont {V.}~\bibnamefont
		{Mukherjee}}, \bibinfo {author} {\bibfnamefont {A.}~\bibnamefont {Dutta}}, \
	and\ \bibinfo {author} {\bibfnamefont {D.}~\bibnamefont {Sen}},\ }\href
{http://stacks.iop.org/1742-5468/2009/i=02/a=P02007} {\bibfield  {journal}
	{\bibinfo  {journal} {J. Stat. Mech. Theor. Exp.}\ }\textbf {\bibinfo
		{volume} {2009}},\ \bibinfo {pages} {P02007} (\bibinfo {year}
	{2009})}\BibitemShut {NoStop}%
\bibitem [{\citenamefont {Divakaran}\ \emph {et~al.}(2008)\citenamefont
	{Divakaran}, \citenamefont {Dutta},\ and\ \citenamefont {Sen}}]{ind3}%
\BibitemOpen
\bibfield  {author} {\bibinfo {author} {\bibfnamefont {U.}~\bibnamefont
		{Divakaran}}, \bibinfo {author} {\bibfnamefont {A.}~\bibnamefont {Dutta}}, \
	and\ \bibinfo {author} {\bibfnamefont {D.}~\bibnamefont {Sen}},\ }\href
{\doibase 10.1103/PhysRevB.78.144301} {\bibfield  {journal} {\bibinfo
		{journal} {Phys. Rev. B}\ }\textbf {\bibinfo {volume} {78}},\ \bibinfo
	{pages} {144301} (\bibinfo {year} {2008})}\BibitemShut {NoStop}%
\bibitem [{\citenamefont {Chowdhury}\ \emph {et~al.}(2010)\citenamefont
	{Chowdhury}, \citenamefont {Divakaran},\ and\ \citenamefont {Dutta}}]{ind4}%
\BibitemOpen
\bibfield  {author} {\bibinfo {author} {\bibfnamefont {D.}~\bibnamefont
		{Chowdhury}}, \bibinfo {author} {\bibfnamefont {U.}~\bibnamefont
		{Divakaran}}, \ and\ \bibinfo {author} {\bibfnamefont {A.}~\bibnamefont
		{Dutta}},\ }\href {\doibase 10.1103/PhysRevE.81.012101} {\bibfield  {journal}
	{\bibinfo  {journal} {Phys. Rev. E}\ }\textbf {\bibinfo {volume} {81}},\
	\bibinfo {pages} {012101} (\bibinfo {year} {2010})}\BibitemShut {NoStop}%
\bibitem [{\citenamefont {Sengupta}\ \emph {et~al.}(2008)\citenamefont
	{Sengupta}, \citenamefont {Sen},\ and\ \citenamefont {Mondal}}]{ind5}%
\BibitemOpen
\bibfield  {author} {\bibinfo {author} {\bibfnamefont {K.}~\bibnamefont
		{Sengupta}}, \bibinfo {author} {\bibfnamefont {D.}~\bibnamefont {Sen}}, \
	and\ \bibinfo {author} {\bibfnamefont {S.}~\bibnamefont {Mondal}},\ }\href
{\doibase 10.1103/PhysRevLett.100.077204} {\bibfield  {journal} {\bibinfo
		{journal} {Phys. Rev. Lett.}\ }\textbf {\bibinfo {volume} {100}},\ \bibinfo
	{pages} {077204} (\bibinfo {year} {2008})}\BibitemShut {NoStop}%
\bibitem [{\citenamefont {Mondal}\ \emph {et~al.}(2008)\citenamefont {Mondal},
	\citenamefont {Sen},\ and\ \citenamefont {Sengupta}}]{ind6}%
\BibitemOpen
\bibfield  {author} {\bibinfo {author} {\bibfnamefont {S.}~\bibnamefont
		{Mondal}}, \bibinfo {author} {\bibfnamefont {D.}~\bibnamefont {Sen}}, \ and\
	\bibinfo {author} {\bibfnamefont {K.}~\bibnamefont {Sengupta}},\ }\href
{\doibase 10.1103/PhysRevB.78.045101} {\bibfield  {journal} {\bibinfo
		{journal} {Phys. Rev. B}\ }\textbf {\bibinfo {volume} {78}},\ \bibinfo
	{pages} {045101} (\bibinfo {year} {2008})}\BibitemShut {NoStop}%
\bibitem [{\citenamefont {Divakaran}\ and\ \citenamefont {Dutta}(2009)}]{ind7}%
\BibitemOpen
\bibfield  {author} {\bibinfo {author} {\bibfnamefont {U.}~\bibnamefont
		{Divakaran}}\ and\ \bibinfo {author} {\bibfnamefont {A.}~\bibnamefont
		{Dutta}},\ }\href {\doibase 10.1103/PhysRevB.79.224408} {\bibfield  {journal}
	{\bibinfo  {journal} {Phys. Rev. B}\ }\textbf {\bibinfo {volume} {79}},\
	\bibinfo {pages} {224408} (\bibinfo {year} {2009})}\BibitemShut {NoStop}%
\bibitem [{\citenamefont {Mukherjee}\ \emph {et~al.}(2008)\citenamefont
	{Mukherjee}, \citenamefont {Dutta},\ and\ \citenamefont {Sen}}]{ind8}%
\BibitemOpen
\bibfield  {author} {\bibinfo {author} {\bibfnamefont {V.}~\bibnamefont
		{Mukherjee}}, \bibinfo {author} {\bibfnamefont {A.}~\bibnamefont {Dutta}}, \
	and\ \bibinfo {author} {\bibfnamefont {D.}~\bibnamefont {Sen}},\ }\href
{\doibase 10.1103/PhysRevB.77.214427} {\bibfield  {journal} {\bibinfo
		{journal} {Phys. Rev. B}\ }\textbf {\bibinfo {volume} {77}},\ \bibinfo
	{pages} {214427} (\bibinfo {year} {2008})}\BibitemShut {NoStop}%
\bibitem [{\citenamefont {Mukherjee}\ and\ \citenamefont {Dutta}(2009)}]{ind9}%
\BibitemOpen
\bibfield  {author} {\bibinfo {author} {\bibfnamefont {V.}~\bibnamefont
		{Mukherjee}}\ and\ \bibinfo {author} {\bibfnamefont {A.}~\bibnamefont
		{Dutta}},\ }\href {http://stacks.iop.org/1742-5468/2009/i=05/a=P05005}
{\bibfield  {journal} {\bibinfo  {journal} {J. Stat. Mech. Theor. Exp.}\
	}\textbf {\bibinfo {volume} {2009}},\ \bibinfo {pages} {P05005} (\bibinfo
	{year} {2009})}\BibitemShut {NoStop}%
\bibitem [{\citenamefont {Divakaran}\ \emph {et~al.}(2010)\citenamefont
	{Divakaran}, \citenamefont {Dutta},\ and\ \citenamefont {Sen}}]{ind10}%
\BibitemOpen
\bibfield  {author} {\bibinfo {author} {\bibfnamefont {U.}~\bibnamefont
		{Divakaran}}, \bibinfo {author} {\bibfnamefont {A.}~\bibnamefont {Dutta}}, \
	and\ \bibinfo {author} {\bibfnamefont {D.}~\bibnamefont {Sen}},\ }\href
{\doibase 10.1103/PhysRevB.81.054306} {\bibfield  {journal} {\bibinfo
		{journal} {Phys. Rev. B}\ }\textbf {\bibinfo {volume} {81}},\ \bibinfo
	{pages} {054306} (\bibinfo {year} {2010})}\BibitemShut {NoStop}%
\bibitem [{\citenamefont {Suzuki}(2009)}]{ind11}%
\BibitemOpen
\bibfield  {author} {\bibinfo {author} {\bibfnamefont {S.}~\bibnamefont
		{Suzuki}},\ }\href {http://stacks.iop.org/1742-5468/2009/i=03/a=P03032}
{\bibfield  {journal} {\bibinfo  {journal} {J. Stat. Mech. Theor. Exp.}\
	}\textbf {\bibinfo {volume} {2009}},\ \bibinfo {pages} {P03032} (\bibinfo
	{year} {2009})}\BibitemShut {NoStop}%
\bibitem [{\citenamefont {Dutta}\ \emph {et~al.}(2010)\citenamefont {Dutta},
	\citenamefont {Singh},\ and\ \citenamefont {Divakaran}}]{ind12}%
\BibitemOpen
\bibfield  {author} {\bibinfo {author} {\bibfnamefont {A.}~\bibnamefont
		{Dutta}}, \bibinfo {author} {\bibfnamefont {R.~R.~P.}\ \bibnamefont {Singh}},
	\ and\ \bibinfo {author} {\bibfnamefont {U.}~\bibnamefont {Divakaran}},\
}\href {http://stacks.iop.org/0295-5075/89/i=6/a=67001} {\bibfield  {journal}
{\bibinfo  {journal} {EPL}\ }\textbf {\bibinfo {volume} {89}},\ \bibinfo
{pages} {67001} (\bibinfo {year} {2010})}\BibitemShut {NoStop}%
\bibitem [{\citenamefont {D\'ora}\ and\ \citenamefont
	{Moessner}(2010)}]{ind13}%
\BibitemOpen
\bibfield  {author} {\bibinfo {author} {\bibfnamefont {B.}~\bibnamefont
		{D\'ora}}\ and\ \bibinfo {author} {\bibfnamefont {R.}~\bibnamefont
		{Moessner}},\ }\href {\doibase 10.1103/PhysRevB.81.165431} {\bibfield
	{journal} {\bibinfo  {journal} {Phys. Rev. B}\ }\textbf {\bibinfo {volume}
		{81}},\ \bibinfo {pages} {165431} (\bibinfo {year} {2010})}\BibitemShut
{NoStop}%
\bibitem [{\citenamefont {Baumann}\ \emph {et~al.}(2011)\citenamefont
	{Baumann}, \citenamefont {Mottl}, \citenamefont {Brennecke},\ and\
	\citenamefont {Esslinger}}]{Esslinger}%
\BibitemOpen
\bibfield  {author} {\bibinfo {author} {\bibfnamefont {K.}~\bibnamefont
		{Baumann}}, \bibinfo {author} {\bibfnamefont {R.}~\bibnamefont {Mottl}},
	\bibinfo {author} {\bibfnamefont {F.}~\bibnamefont {Brennecke}}, \ and\
	\bibinfo {author} {\bibfnamefont {T.}~\bibnamefont {Esslinger}},\ }\href
{\doibase 10.1103/PhysRevLett.107.140402} {\bibfield  {journal} {\bibinfo
		{journal} {Phys. Rev. Lett.}\ }\textbf {\bibinfo {volume} {107}},\ \bibinfo
	{pages} {140402} (\bibinfo {year} {2011})}\BibitemShut {NoStop}%
\bibitem [{\citenamefont {Chen}\ \emph {et~al.}(2011)\citenamefont {Chen},
	\citenamefont {White}, \citenamefont {Borries},\ and\ \citenamefont
	{DeMarco}}]{deMarcoclean}%
\BibitemOpen
\bibfield  {author} {\bibinfo {author} {\bibfnamefont {D.}~\bibnamefont
		{Chen}}, \bibinfo {author} {\bibfnamefont {M.}~\bibnamefont {White}},
	\bibinfo {author} {\bibfnamefont {C.}~\bibnamefont {Borries}}, \ and\
	\bibinfo {author} {\bibfnamefont {B.}~\bibnamefont {DeMarco}},\ }\href
{\doibase 10.1103/PhysRevLett.106.235304} {\bibfield  {journal} {\bibinfo
		{journal} {Phys. Rev. Lett.}\ }\textbf {\bibinfo {volume} {106}},\ \bibinfo
	{pages} {235304} (\bibinfo {year} {2011})}\BibitemShut {NoStop}%
\bibitem [{\citenamefont {Braun}\ \emph {et~al.}(2015)\citenamefont {Braun},
	\citenamefont {Friesdorf}, \citenamefont {Hodgman}, \citenamefont
	{Schreiber}, \citenamefont {Ronzheimer}, \citenamefont {Riera}, \citenamefont
	{del Rey}, \citenamefont {Bloch}, \citenamefont {Eisert},\ and\ \citenamefont
	{Schneider}}]{Schaetz}%
\BibitemOpen
\bibfield  {author} {\bibinfo {author} {\bibfnamefont {S.}~\bibnamefont
		{Braun}}, \bibinfo {author} {\bibfnamefont {M.}~\bibnamefont {Friesdorf}},
	\bibinfo {author} {\bibfnamefont {S.~S.}\ \bibnamefont {Hodgman}}, \bibinfo
	{author} {\bibfnamefont {M.}~\bibnamefont {Schreiber}}, \bibinfo {author}
	{\bibfnamefont {J.~P.}\ \bibnamefont {Ronzheimer}}, \bibinfo {author}
	{\bibfnamefont {A.}~\bibnamefont {Riera}}, \bibinfo {author} {\bibfnamefont
		{M.}~\bibnamefont {del Rey}}, \bibinfo {author} {\bibfnamefont
		{I.}~\bibnamefont {Bloch}}, \bibinfo {author} {\bibfnamefont
		{J.}~\bibnamefont {Eisert}}, \ and\ \bibinfo {author} {\bibfnamefont
		{U.}~\bibnamefont {Schneider}},\ }\href {\doibase 10.1073/pnas.1408861112}
{\bibfield  {journal} {\bibinfo  {journal} {PNAS}\ }\textbf {\bibinfo
		{volume} {112}},\ \bibinfo {pages} {3641} (\bibinfo {year}
	{2015})}\BibitemShut {NoStop}%
\bibitem [{\citenamefont {Meldgin}\ \emph {et~al.}(2016)\citenamefont
	{Meldgin}, \citenamefont {Ray}, \citenamefont {Russ}, \citenamefont {Chen},
	\citenamefont {Ceperley},\ and\ \citenamefont {DeMarco}}]{deMarcodisorder}%
\BibitemOpen
\bibfield  {author} {\bibinfo {author} {\bibfnamefont {C.}~\bibnamefont
		{Meldgin}}, \bibinfo {author} {\bibfnamefont {U.}~\bibnamefont {Ray}},
	\bibinfo {author} {\bibfnamefont {P.}~\bibnamefont {Russ}}, \bibinfo {author}
	{\bibfnamefont {D.}~\bibnamefont {Chen}}, \bibinfo {author} {\bibfnamefont
		{D.~M.}\ \bibnamefont {Ceperley}}, \ and\ \bibinfo {author} {\bibfnamefont
		{B.}~\bibnamefont {DeMarco}},\ }\href {http://dx.doi.org/10.1038/nphys3695}
{\bibfield  {journal} {\bibinfo  {journal} {Nat Phys}\ }\textbf {\bibinfo
		{volume} {12}},\ \bibinfo {pages} {646} (\bibinfo {year} {2016})},\ \bibinfo
{note} {letter}\BibitemShut {NoStop}%
\bibitem [{\citenamefont {Cui}\ \emph {et~al.}(2016)\citenamefont {Cui},
	\citenamefont {Huang}, \citenamefont {Wang}, \citenamefont {Cao},
	\citenamefont {Wang}, \citenamefont {Lv}, \citenamefont {Luo}, \citenamefont
	{del Campo}, \citenamefont {Han}, \citenamefont {Li},\ and\ \citenamefont
	{Guo}}]{chinskiLZ}%
\BibitemOpen
\bibfield  {author} {\bibinfo {author} {\bibfnamefont {J.-M.}\ \bibnamefont
		{Cui}}, \bibinfo {author} {\bibfnamefont {Y.-F.}\ \bibnamefont {Huang}},
	\bibinfo {author} {\bibfnamefont {Z.}~\bibnamefont {Wang}}, \bibinfo {author}
	{\bibfnamefont {D.-Y.}\ \bibnamefont {Cao}}, \bibinfo {author} {\bibfnamefont
		{J.}~\bibnamefont {Wang}}, \bibinfo {author} {\bibfnamefont {W.-M.}\
		\bibnamefont {Lv}}, \bibinfo {author} {\bibfnamefont {L.}~\bibnamefont
		{Luo}}, \bibinfo {author} {\bibfnamefont {A.}~\bibnamefont {del Campo}},
	\bibinfo {author} {\bibfnamefont {Y.-J.}\ \bibnamefont {Han}}, \bibinfo
	{author} {\bibfnamefont {C.-F.}\ \bibnamefont {Li}}, \ and\ \bibinfo {author}
	{\bibfnamefont {G.-C.}\ \bibnamefont {Guo}},\ }\href
{http://dx.doi.org/10.1038/srep33381} {\bibfield  {journal} {\bibinfo
		{journal} {Scic Rep.}\ }\textbf {\bibinfo {volume} {6}},\ \bibinfo {pages}
	{33381} (\bibinfo {year} {2016})}\BibitemShut {NoStop}%
\bibitem [{\citenamefont {Kolodrubetz}\ \emph {et~al.}(2012)\citenamefont
	{Kolodrubetz}, \citenamefont {Clark},\ and\ \citenamefont
	{Huse}}]{Kolodrubetz}%
\BibitemOpen
\bibfield  {author} {\bibinfo {author} {\bibfnamefont {M.}~\bibnamefont
		{Kolodrubetz}}, \bibinfo {author} {\bibfnamefont {B.~K.}\ \bibnamefont
		{Clark}}, \ and\ \bibinfo {author} {\bibfnamefont {D.~A.}\ \bibnamefont
		{Huse}},\ }\href {\doibase 10.1103/PhysRevLett.109.015701} {\bibfield
	{journal} {\bibinfo  {journal} {Phys. Rev. Lett.}\ }\textbf {\bibinfo
		{volume} {109}},\ \bibinfo {pages} {015701} (\bibinfo {year}
	{2012})}\BibitemShut {NoStop}%
\bibitem [{\citenamefont {Chandran}\ \emph {et~al.}(2012)\citenamefont
	{Chandran}, \citenamefont {Erez}, \citenamefont {Gubser},\ and\ \citenamefont
	{Sondhi}}]{princeton}%
\BibitemOpen
\bibfield  {author} {\bibinfo {author} {\bibfnamefont {A.}~\bibnamefont
		{Chandran}}, \bibinfo {author} {\bibfnamefont {A.}~\bibnamefont {Erez}},
	\bibinfo {author} {\bibfnamefont {S.~S.}\ \bibnamefont {Gubser}}, \ and\
	\bibinfo {author} {\bibfnamefont {S.~L.}\ \bibnamefont {Sondhi}},\ }\href
{\doibase 10.1103/PhysRevB.86.064304} {\bibfield  {journal} {\bibinfo
		{journal} {Phys. Rev. B}\ }\textbf {\bibinfo {volume} {86}},\ \bibinfo
	{pages} {064304} (\bibinfo {year} {2012})}\BibitemShut {NoStop}%
\bibitem [{\citenamefont {Deng}\ \emph {et~al.}(2008)\citenamefont {Deng},
	\citenamefont {Ortiz},\ and\ \citenamefont {Viola}}]{ViolaOrtiz}%
\BibitemOpen
\bibfield  {author} {\bibinfo {author} {\bibfnamefont {S.}~\bibnamefont
		{Deng}}, \bibinfo {author} {\bibfnamefont {G.}~\bibnamefont {Ortiz}}, \ and\
	\bibinfo {author} {\bibfnamefont {L.}~\bibnamefont {Viola}},\ }\href
{http://stacks.iop.org/0295-5075/84/i=6/a=67008} {\bibfield  {journal}
	{\bibinfo  {journal} {EPL}\ }\textbf {\bibinfo {volume} {84}},\ \bibinfo
	{pages} {67008} (\bibinfo {year} {2008})}\BibitemShut {NoStop}%
\bibitem [{\citenamefont {Damski}\ and\ \citenamefont
	{Zurek}(2009)}]{DamskiZurek}%
\BibitemOpen
\bibfield  {author} {\bibinfo {author} {\bibfnamefont {B.}~\bibnamefont
		{Damski}}\ and\ \bibinfo {author} {\bibfnamefont {W.~H.}\ \bibnamefont
		{Zurek}},\ }\href {http://stacks.iop.org/1367-2630/11/i=6/a=063014}
{\bibfield  {journal} {\bibinfo  {journal} {New J. Phys.}\ }\textbf {\bibinfo
		{volume} {11}},\ \bibinfo {pages} {063014} (\bibinfo {year}
	{2009})}\BibitemShut {NoStop}%
\bibitem [{\citenamefont {Dziarmaga}\ and\ \citenamefont
	{Rams}(2010)}]{DzRams}%
\BibitemOpen
\bibfield  {author} {\bibinfo {author} {\bibfnamefont {J.}~\bibnamefont
		{Dziarmaga}}\ and\ \bibinfo {author} {\bibfnamefont {M.~M.}\ \bibnamefont
		{Rams}},\ }\href {http://stacks.iop.org/1367-2630/12/i=5/a=055007} {\bibfield
	{journal} {\bibinfo  {journal} {New J. Phys.}\ }\textbf {\bibinfo {volume}
		{12}},\ \bibinfo {pages} {055007} (\bibinfo {year} {2010})}\BibitemShut
{NoStop}%
\bibitem [{\citenamefont {Dziarmaga}\ and\ \citenamefont {Zurek}(2014)}]{KZSR}%
\BibitemOpen
\bibfield  {author} {\bibinfo {author} {\bibfnamefont {J.}~\bibnamefont
		{Dziarmaga}}\ and\ \bibinfo {author} {\bibfnamefont {W.~H.}\ \bibnamefont
		{Zurek}},\ }\href {http://dx.doi.org/10.1038/srep05950} {\bibfield  {journal}
	{\bibinfo  {journal} {Sci. Rep.}\ }\textbf {\bibinfo {volume} {4}},\ \bibinfo
	{pages} {5950} (\bibinfo {year} {2014})}\BibitemShut {NoStop}%
\bibitem [{\citenamefont {Kosterlitz}\ and\ \citenamefont
	{Thouless}(1973)}]{KT1}%
\BibitemOpen
\bibfield  {author} {\bibinfo {author} {\bibfnamefont {J.~M.}\ \bibnamefont
		{Kosterlitz}}\ and\ \bibinfo {author} {\bibfnamefont {D.~J.}\ \bibnamefont
		{Thouless}},\ }\href {http://stacks.iop.org/0022-3719/6/i=7/a=010} {\bibfield
	{journal} {\bibinfo  {journal} {J. Phys. C}\ }\textbf {\bibinfo {volume}
		{6}},\ \bibinfo {pages} {1181} (\bibinfo {year} {1973})}\BibitemShut
{NoStop}%
\bibitem [{\citenamefont {Resnick}\ \emph {et~al.}(1981)\citenamefont
	{Resnick}, \citenamefont {Garland}, \citenamefont {Boyd}, \citenamefont
	{Shoemaker},\ and\ \citenamefont {Newrock}}]{KT2}%
\BibitemOpen
\bibfield  {author} {\bibinfo {author} {\bibfnamefont {D.~J.}\ \bibnamefont
		{Resnick}}, \bibinfo {author} {\bibfnamefont {J.~C.}\ \bibnamefont
		{Garland}}, \bibinfo {author} {\bibfnamefont {J.~T.}\ \bibnamefont {Boyd}},
	\bibinfo {author} {\bibfnamefont {S.}~\bibnamefont {Shoemaker}}, \ and\
	\bibinfo {author} {\bibfnamefont {R.~S.}\ \bibnamefont {Newrock}},\ }\href
{\doibase 10.1103/PhysRevLett.47.1542} {\bibfield  {journal} {\bibinfo
		{journal} {Phys. Rev. Lett.}\ }\textbf {\bibinfo {volume} {47}},\ \bibinfo
	{pages} {1542} (\bibinfo {year} {1981})}\BibitemShut {NoStop}%
\bibitem [{\citenamefont {Zoran}\ \emph {et~al.}(2006)\citenamefont {Zoran},
	\citenamefont {Peter}, \citenamefont {Marc}, \citenamefont {Baptiste},\ and\
	\citenamefont {Dalibard}}]{KT3}%
\BibitemOpen
\bibfield  {author} {\bibinfo {author} {\bibfnamefont {H.}~\bibnamefont
		{Zoran}}, \bibinfo {author} {\bibfnamefont {K.}~\bibnamefont {Peter}},
	\bibinfo {author} {\bibfnamefont {C.}~\bibnamefont {Marc}}, \bibinfo {author}
	{\bibfnamefont {B.}~\bibnamefont {Baptiste}}, \ and\ \bibinfo {author}
	{\bibfnamefont {J.}~\bibnamefont {Dalibard}},\ }\href {\doibase
	10.1038/nature04851} {\bibfield  {journal} {\bibinfo  {journal} {Nature}\
	}\textbf {\bibinfo {volume} {441}},\ \bibinfo {pages} {1118} (\bibinfo {year}
	{2006})}\BibitemShut {NoStop}%
\bibitem [{\citenamefont {Girvin}(2000)}]{Girvin}%
\BibitemOpen
\bibfield  {author} {\bibinfo {author} {\bibfnamefont {S.~M.}\ \bibnamefont
		{Girvin}},\ }\href@noop {} {\enquote {\bibinfo {title} {The
			{K}osterlitz-{T}houless transition},}\ } (\bibinfo {year} {unpublished
	lecture notes Boulder, 2000})\BibitemShut {NoStop}%
\bibitem [{\citenamefont {Corless}(1993)}]{Corless93}%
\BibitemOpen
\bibfield  {author} {\bibinfo {author} {\bibfnamefont {R.}~\bibnamefont
		{Corless}},\ }\href {https://books.google.com/books?id=TumZHAAACAAJ} {\emph
	{\bibinfo {title} {On the Lambert W Function}}},\ Research report\ (\bibinfo
{publisher} {University of Waterloo, Computer Science Department},\ \bibinfo
{year} {1993})\BibitemShut {NoStop}%
\bibitem [{\citenamefont {Jeli{\'c}}\ and\ \citenamefont
	{Cugliandolo}(2011)}]{BKTclass}%
\BibitemOpen
\bibfield  {author} {\bibinfo {author} {\bibfnamefont {A.}~\bibnamefont
		{Jeli{\'c}}}\ and\ \bibinfo {author} {\bibfnamefont {L.~F.}\ \bibnamefont
		{Cugliandolo}},\ }\href {http://stacks.iop.org/1742-5468/2011/i=02/a=P02032}
{\bibfield  {journal} {\bibinfo  {journal} {J. Stat. Mech. Theor. Exp.}\
	}\textbf {\bibinfo {volume} {2011}},\ \bibinfo {pages} {P02032} (\bibinfo
	{year} {2011})}\BibitemShut {NoStop}%
\bibitem [{\citenamefont {Deutschl\"{a}nder}\ \emph {et~al.}(2015)\citenamefont
	{Deutschl\"{a}nder}, \citenamefont {Dillmann}, \citenamefont {Maret},\ and\
	\citenamefont {Keim}}]{Deutschlander15}%
\BibitemOpen
\bibfield  {author} {\bibinfo {author} {\bibfnamefont {S.}~\bibnamefont
		{Deutschl\"{a}nder}}, \bibinfo {author} {\bibfnamefont {P.}~\bibnamefont
		{Dillmann}}, \bibinfo {author} {\bibfnamefont {G.}~\bibnamefont {Maret}}, \
	and\ \bibinfo {author} {\bibfnamefont {P.}~\bibnamefont {Keim}},\ }\href
{\doibase 10.1073/pnas.1500763112} {\bibfield  {journal} {\bibinfo  {journal}
		{PNAS}\ }\textbf {\bibinfo {volume} {112}},\ \bibinfo {pages} {6925}
	(\bibinfo {year} {2015})}\BibitemShut {NoStop}%
\bibitem [{\citenamefont {Wall}\ and\ \citenamefont {Carr}(2012)}]{Wall12}%
\BibitemOpen
\bibfield  {author} {\bibinfo {author} {\bibfnamefont {M.~L.}\ \bibnamefont
		{Wall}}\ and\ \bibinfo {author} {\bibfnamefont {L.~D.}\ \bibnamefont
		{Carr}},\ }\href {\doibase 10.1088/1367-2630/14/12/125015} {\bibfield
	{journal} {\bibinfo  {journal} {New J. Phys.}\ }\textbf {\bibinfo {volume}
		{14}},\ \bibinfo {pages} {125015} (\bibinfo {year} {2012})}\BibitemShut
{NoStop}%
\bibitem [{\citenamefont {Cucchietti}\ \emph {et~al.}(2007)\citenamefont
	{Cucchietti}, \citenamefont {Damski}, \citenamefont {Dziarmaga},\ and\
	\citenamefont {Zurek}}]{DHBog}%
\BibitemOpen
\bibfield  {author} {\bibinfo {author} {\bibfnamefont {F.~M.}\ \bibnamefont
		{Cucchietti}}, \bibinfo {author} {\bibfnamefont {B.}~\bibnamefont {Damski}},
	\bibinfo {author} {\bibfnamefont {J.}~\bibnamefont {Dziarmaga}}, \ and\
	\bibinfo {author} {\bibfnamefont {W.~H.}\ \bibnamefont {Zurek}},\ }\href
{\doibase 10.1103/PhysRevA.75.023603} {\bibfield  {journal} {\bibinfo
		{journal} {Phys. Rev. A}\ }\textbf {\bibinfo {volume} {75}},\ \bibinfo
	{pages} {023603} (\bibinfo {year} {2007})}\BibitemShut {NoStop}%
\bibitem [{\citenamefont {Elstner}\ and\ \citenamefont {Monien}(1999)}]{Jc}%
\BibitemOpen
\bibfield  {author} {\bibinfo {author} {\bibfnamefont {N.}~\bibnamefont
		{Elstner}}\ and\ \bibinfo {author} {\bibfnamefont {H.}~\bibnamefont
		{Monien}},\ }\href@noop {} {\enquote {\bibinfo {title} {A numerical exact
			solution of the {B}ose-{H}ubbard model},}\ } (\bibinfo {year} {1999}),\
\Eprint {http://arxiv.org/abs/cond-mat/9905367} {arXiv:cond-mat/9905367}
\BibitemShut {NoStop}%
\bibitem [{\citenamefont {Gritsev}\ and\ \citenamefont
	{Polkovnikov}(2010)}]{Gritsev10}%
\BibitemOpen
\bibfield  {author} {\bibinfo {author} {\bibfnamefont {V.}~\bibnamefont
		{Gritsev}}\ and\ \bibinfo {author} {\bibfnamefont {A.}~\bibnamefont
		{Polkovnikov}},\ }\href@noop {} {\enquote {\bibinfo {title} {Universal
			dynamics near quantum critical points},}\ } (\bibinfo {year} {2010}),\
\Eprint {http://arxiv.org/abs/0910.3692v3} {arXiv:0910.3692v3} \BibitemShut
{NoStop}%
\bibitem [{\citenamefont {Binder}\ \emph {et~al.}(2015)\citenamefont {Binder},
	\citenamefont {Vinjanampathy}, \citenamefont {Modi},\ and\ \citenamefont
	{Goold}}]{Binder15}%
\BibitemOpen
\bibfield  {author} {\bibinfo {author} {\bibfnamefont {F.}~\bibnamefont
		{Binder}}, \bibinfo {author} {\bibfnamefont {S.}~\bibnamefont
		{Vinjanampathy}}, \bibinfo {author} {\bibfnamefont {K.}~\bibnamefont {Modi}},
	\ and\ \bibinfo {author} {\bibfnamefont {J.}~\bibnamefont {Goold}},\ }\href
{\doibase 10.1103/PhysRevE.91.032119} {\bibfield  {journal} {\bibinfo
		{journal} {Phys. Rev. E}\ }\textbf {\bibinfo {volume} {91}},\ \bibinfo
	{pages} {032119} (\bibinfo {year} {2015})}\BibitemShut {NoStop}%
\bibitem [{\citenamefont {Gardas}\ and\ \citenamefont
	{Deffner}(2015)}]{Gardas15}%
\BibitemOpen
\bibfield  {author} {\bibinfo {author} {\bibfnamefont {B.}~\bibnamefont
		{Gardas}}\ and\ \bibinfo {author} {\bibfnamefont {S.}~\bibnamefont
		{Deffner}},\ }\href {\doibase 10.1103/PhysRevE.92.042126} {\bibfield
	{journal} {\bibinfo  {journal} {Phys. Rev. E}\ }\textbf {\bibinfo {volume}
		{92}},\ \bibinfo {pages} {042126} (\bibinfo {year} {2015})}\BibitemShut
{NoStop}%
\bibitem [{\citenamefont {Dziarmaga}\ and\ \citenamefont
	{Tylutki}(2011)}]{TylutkiLutt}%
\BibitemOpen
\bibfield  {author} {\bibinfo {author} {\bibfnamefont {J.}~\bibnamefont
		{Dziarmaga}}\ and\ \bibinfo {author} {\bibfnamefont {M.}~\bibnamefont
		{Tylutki}},\ }\href {\doibase 10.1103/PhysRevB.84.214522} {\bibfield
	{journal} {\bibinfo  {journal} {Phys. Rev. B}\ }\textbf {\bibinfo {volume}
		{84}},\ \bibinfo {pages} {214522} (\bibinfo {year} {2011})}\BibitemShut
{NoStop}%
\end{thebibliography}
%
\end{document}